\documentclass[12pt]{iopart}

\usepackage{graphicx}
\usepackage{color}

\newcommand{\nuc}[2]{$^{#1}$#2}
\newcommand{\nucm}[2]{$^{#1m}$#2}
\def\F{{\cal F}}
\def\GV{G_{\mbox{\tiny V}}}
\def\GF{G_{\mbox{\tiny F}}}
\def\DRV{\Delta_{\mbox{\tiny R}}^{\mbox{\tiny V}}}
\def\be {\begin{equation}}
\def\ee {\end{equation}}
\def\bea {\begin{eqnarray}}
\def\eea {\end{eqnarray}}
\newcommand{\sfrac}[2]{\mbox{\small{$\frac{#1}{#2}$}}}
  
\begin{document}

\review{The measurement and interpretation of superallowed $0^+$$\rightarrow 0^+$\,nuclear $\beta$ decay}

\author{J C Hardy and I S Towner}

\address{Cyclotron Institute, Texas A\&M University, College Station, TX 77843-3366, U.S.A.}
\eads{\mailto{hardy@comp.tamu.edu}, \mailto{towner@comp.tamu.edu}}
\begin{abstract}
Measurements of the decay strength of superallowed $0^+$$\rightarrow 0^+$\,nuclear $\beta$ transitions shed light on the fundamental
properties of weak interactions.  Because of their impact, such measurements were first reported 60 years ago in the early 1950s and
have continued unabated ever since, always taking advantage of improvements in experimental techniques to achieve ever higher precision. 
The results helped first to shape the Electroweak Standard Model but more recently have evolved into sensitively testing that model's
predictions.  Today they provide the most demanding test of vector-current conservation and of the unitarity of the Cabibbo-Kobayashi-Maskawa
matrix.  Here, we review the experimental and theoretical methods
that have been, and are being, used to characterize superallowed $0^+$$\rightarrow 0^+$\,$\beta$ transitions and to extract fundamentally
important parameters from their analysis.
\end{abstract}


\section{\label{intro} Introduction}

In 1953, Sherr and Gerhart published a paper \cite{Sh53} on ``Experimental evidence for the Fermi interaction in the $\beta$ decay of $^{14}$O
and $^{10}$C."  It was less than five years since Sherr had first discovered these two nuclei \cite{Sh49}, yet
already the two authors were using the decays to probe for the first time the fundamental nature of $\beta$ decay.  They were able to identify
superallowed transitions in both decays -- they called them ``allowed favoured transitions" -- and recognized that the Fermi theory of $\beta$ decay
predicted that the comparative half-lives, or $ft$ values, for the two transitions should be the same, a prediction they could test.  The $ft$ value
for a transition depends on the energy released by the transition as well as its branching ratio and the half-life of the initial
state.  Using degraders to determine the energy of the emitted positrons, a NaI(Tl) scintillation spectrometer to establish the branching ratios and
Geiger counters to measure the half lives, Sherr and Gerhard were able to conclude that the two $ft$ values were indeed the same, albeit with
large error bars: They obtained 3300$\pm$750 s for the superallowed $ft$ value of $^{14}$O and 5900$\pm$2400 s for that of $^{10}$C.

By 1960, when Bardin {\it et al.}~reported a much improved result for $^{14}$O \cite{Ba60}, experimental techniques had advanced considerably.  The
decay energy was no longer dependent on positron range measurements, but rather on measured Q values for the reactions $^{12}$C($^3$He,$n$)$^{14}$O and
$^{12}$C($^3$He,$p$)$^{14}$N*, which populated the superallowed transition's parent and daughter states, respectively, from a common target.  Also,
the branching ratio could be corrected for the weak non-superallowed $\beta$ branch to the ground state of $^{14}$N, which had by then been measured
with the help of a magnetic lens spectrometer \cite{Sh55}.  They reported an $ft$ value of 3060$\pm$13 s, which for its time is a remarkably precise
result and stands only 1.4 standard deviations away from the currently accepted value of 3042.3$\pm$2.7 s.

The precision obtained by Bardin {\it et al.}~was sufficient for them to compare the vector coupling constant $\GV$, derived from their $ft$-value,
with $\GF$, the weak-interaction constant derived from the purely leptonic decay of the muon. The Conserved Vector Current (CVC) hypothesis had been
proposed by Feynman and Gell-Mann two years earlier \cite{Fe58} but the role of strangeness was not yet understood, so the authors expected that $\GV$
should equal $\GF$.  Although their measurement actually showed a small difference between the two, the authors noted that radiative and Coulomb corrections
could account for the discrepancy and concluded that it was too soon to tell if this constituted a failure of the universality hypothesis.

The discrepancy persisted, however, and was joined by other discrepancies observed between weak decays that changed strangeness $S$ and those that
did not, the most glaring example being the decays of the $K^+$ ($S$=1) and the $\pi^+$ ($S$=0) mesons to the same final state, $\mu^+$ + $\nu$.
All these apparent conflicts with vector-current universality were resolved by Cabibbo in 1963 \cite{Ca63} when he recognized that the universality
of the weak interaction was manifest only if one considered the total strength of both the strangeness non-changing and the strangeness changing decays.
In modern terminology we would say that he was the first to realize that there is mixing between the first two generations of quarks, and to express
that mixing in terms of a unitary rotation.

As it turned out, that was only part of the story.  A year after Cabibbo introduced his rotation angle, another symmetry -- that of CP -- was observed to
be violated in the weak decay of the long-lived neutral kaon.  This result, which remained a puzzle for nearly a decade, ultimately led Kobayashi and Maskawa
\cite{Ko73} in 1973 to postulate the existence of a third generation of quarks -- subsequently confirmed by experiment -- and to replace Cabibbo's single
rotation angle $\theta_C$, which was in effect a 2$\times$2 rotation matrix, by the now familiar 3$\times$3 unitary rotation matrix referred to as the
Cabibbo-Kobayashi-Maskawa, or CKM, matrix.  It became one of the pillars of the Standard Model.

In the conclusion to his 1963 paper, Cabibbo remarked that $\GV$ should no longer be expected to be equal to $\GF$ but rather to $\GF$cos\,$\theta_C$.
However, he noted that although this change was ``in the right direction to eliminate the discrepancy between $^{14}$O and muon lifetimes," in fact the
correction was ``too large, leaving about 2\% to be explained" \cite{Ca63}.  That observation was enough to stimulate a great deal of activity in superallowed $\beta$
decay during the decade between Cabibbo's insight and that of Kobayashi and Maskawa.  By 1973, the number of well-measured superallowed transitions had
grown from one to seven; the 2\% discrepancy was being explained in terms of radiative corrections and charge-dependent nuclear corrections \cite{To73}; and
plausible values for the Cabibbo angle had been extracted from the results.  Once the CKM matrix took center stage though, it was not long before
the focus of superallowed $\beta$ decay had shifted to a determination of the upper left element of the CKM matrix, $V_{ud}$, the value of which is in fact closely
related to $\cos{\theta_C}$.   

Today, 40 years on, although the Standard Model is by now well established, the limits of its applicability are still being probed aggressively.  One such experimental
probe is to test the unitarity of the CKM matrix.  Although the Standard Model does not prescribe values for the nine elements of the matrix -- they must all be
determined from experiment -- it does require the matrix itself to be unitary.  Superallowed $\beta$ decay is currently the source of the most precise value for $V_{ud}$
\cite{Ha09,PDG12}, which is the largest element in the matrix, and is therefore a crucial contributor to the most sensitive available test of CKM unitarity: the sum of
squares of the three top-row elements \cite{TH10}. The current value for that unitarity sum is 1.00008$\pm$0.00056 \cite{Ha13}, a remarkably precise result that
agrees with unitarity and significantly limits the scope for new physics beyond the Standard Model.  However, even this precision, $\pm$0.06\%, will likely be improved
before long by upgraded measurements of superallowed transitions, which are already underway.

In what follows we review particularly the experimental methods that have been, and are being, used to characterize superallowed $0^+$$\rightarrow 0^+$\,$\beta$
transitions.  Although these are all nuclear-physics measurements, the extraordinary demands for precision in the Standard-Model test have motivated the development
of highly refined techniques not commonly employed in other nuclear-physics applications.  In addition, since theory also plays a vital role in the extraction of
$V_{ud}$, we also outline the methods used for calculating the small radiative and isospin-symmetry-breaking corrections that must be applied to the experimental data.

\section{\label{exp} Experiments, past and present} 

Since the axial current cannot contribute to transitions between spin-0 states, superallowed $0^+$$\rightarrow 0^+$ $\beta$ decay between $T$=1 analogue states depends
uniquely on the vector part of the weak interaction.   As already noted, the CVC principle indicates that the experimental $ft$ value for such a transition should be
related to the vector coupling constant, $\GV$, which must be common to all nuclear vector transitions.  In turn, $\GV$ itself is related to the fundamental
weak-interaction coupling constant, $\GF$, via the relation
\be
\GV = \GF V_{ud},
\label{GVtoVud}
\ee
where $\GF /(\hbar c )^3 = (1.1663787 \pm 0.0000006) \times 10^{-5}$ GeV$^{-2}$, as obtained from the measured muon lifetime \cite{PDG12}

In practice, the expression for $ft$
includes several small correction terms.  It is convenient to combine some of these terms with the $ft$ value and define a ``corrected" $\F t$
value.  Thus, we write \cite{Ha09}
\be
\F t \equiv ft (1 + \delta_R^{\prime}) (1 + \delta_{NS} - \delta_C ) = \frac{K}{2 \GV^2 
(1 + \DRV )}~,
\label{Ftconst}
\ee
where $K/(\hbar c )^6 = 2 \pi^3 \hbar \ln 2 / (m_e c^2)^5 = 8120.2787(11) \times 10^{-10}$ GeV$^{-4}$s, $\delta_C$ is the isospin-symmetry-breaking
correction, and $\DRV$ is the transition-independent part of the radiative correction.  The terms $\delta_R^{\prime}$ and $\delta_{NS}$ comprise the
transition-dependent part of the radiative correction, the former being a function only of the electron's energy and the $Z$ of the daughter nucleus,
while the latter, like $\delta_C$, depends in its evaluation on the details of the nuclear structure of the parent and daughter states.  All these
correction terms are of order 1\% or less, with uncertainties at least an order of magnitude smaller than that, so equation~(\ref{Ftconst})
provides an experimental method for determining $\GV$ -- and thus $V_{ud}$ -- to better than a part in a thousand.

Experimentally, the $ft$ value that characterizes a superallowed transition -- or any $\beta$ transition for that matter -- is determined from three
measured quantities: the total transition energy, $Q_{EC}$, the half-life, $t_{1/2}$, of the parent state, and the branching ratio, $R$, for the
particular transition of interest.  The $Q_{EC}$-value is required to determine the phase-space integral, $f$, while the half-life and branching
ratio combine to yield the partial half-life, $t$.  Since the $ft$ value incorporates three experimental quantities, each one of those quantities must
be measured to substantially better than 0.1\% precision in order to achieve that precision on the combination.  This is particularly true for
$Q_{EC}$ since it enters to the fifth power in the calculation of $f$.

\begin{figure}[t]
\hspace{2.3cm}
 \includegraphics[width=10cm]{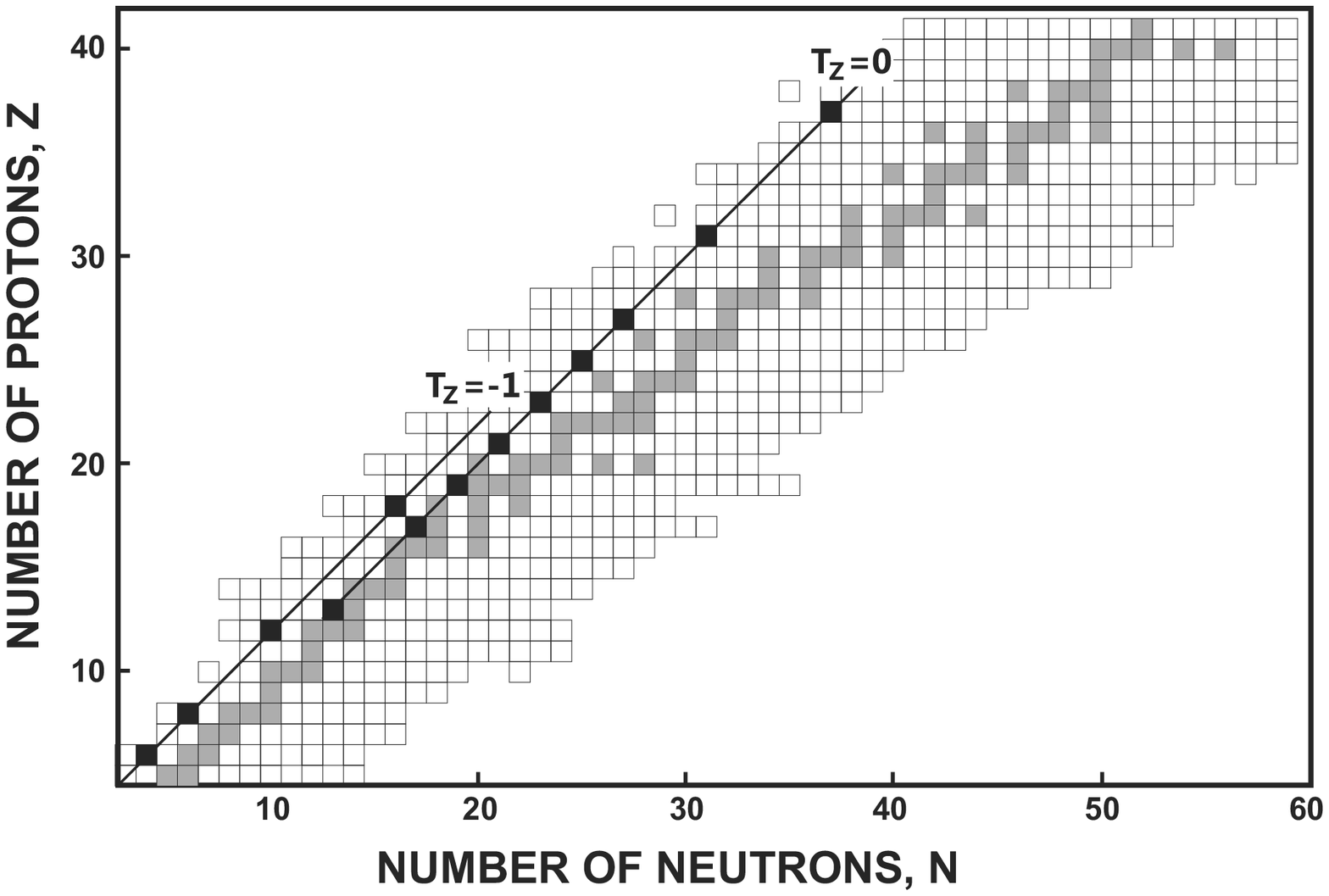}
 \caption{\label{fig:chart}  Partial chart of nuclides showing parents of the precisely measured superallowed transitions as solid black squares.  The
stable nuclei appear as grey squares.  The two labeled diagonal lines mark the loci of the $T_Z$\,=\,$-1$ and $T_Z$\,=\,0 parents.  From left to right,
the former are \nuc{10}{C}, \nuc{14}{O}, \nuc{22}{Mg} and \nuc{34}{Ar}; and the latter are \nucm{26}{Al}, \nuc{34}{Cl}, \nucm{38}{K}, \nuc{42}{Sc},
\nuc{46}{V}, \nuc{50}{Mn}, \nuc{54}{Co}, \nuc{62}{Ga} and \nuc{74}{Rb}.}
\end{figure}

To date, the $ft$ values for ten $0^+$$\rightarrow 0^+$ transitions -- with parents \nuc{14}{O}, \nucm{26}{Al}, \nuc{34}{Cl}, \nucm{38}{K},
\nuc{42}{Sc}, \nuc{46}{V}, \nuc{50}{Mn}, \nuc{54}{Co}, \nuc{62}{Ga} and \nuc{74}{Rb} -- are known to 0.1\% relative precision or better; and
three more -- \nuc{10}{C}, \nuc{22}{Mg} and \nuc{34}{Ar} -- are known to $<$0.3\%.  The 13 cases are shown on the chart of nuclides in Fig.\,\ref{fig:chart}.
How has this level of precision been achieved in all these cases for the combination of three experimental quantities?  In answering this question in the
following sections, we frequently reference examples of various measurement techniques.  A complete referenced list of {\it all} measurements that
currently contribute to world data for superallowed $0^+$$\rightarrow 0^+$ decays appears in the 2009 survey by Hardy and Towner \cite{Ha09}.

\subsection{\label{QEC} $Q_{EC}$-value measurements}

Already in 1960, Barden {\it et al.}~\cite{Ba60} appreciated that a determination of the end-point energy from a measured $\beta$ spectrum, even if
that spectrum were obtained with a magnetic spectrometer, could not possibly match the precision possible with a nuclear reaction.  From that time on,
until the advent of on-line Penning traps less than a decade ago, nuclear reactions were the only method used to determine precise $Q_{EC}$ values.  The
favoured ones were ($p,n$) and ($^3$He,$t$) reactions on the $\beta$-decay daughter nuclei, which are stable for all $0^+$$\rightarrow 0^+$
superallowed decays that have been studied until recently.  Since these reactions connect the same nuclei as the corresponding $\beta$ decay, their
$Q$ values are directly related to the $\beta$-decay $Q_{EC}$ value.

\begin{figure}[t]
\hspace{2.3cm}
 \includegraphics[width=13cm]{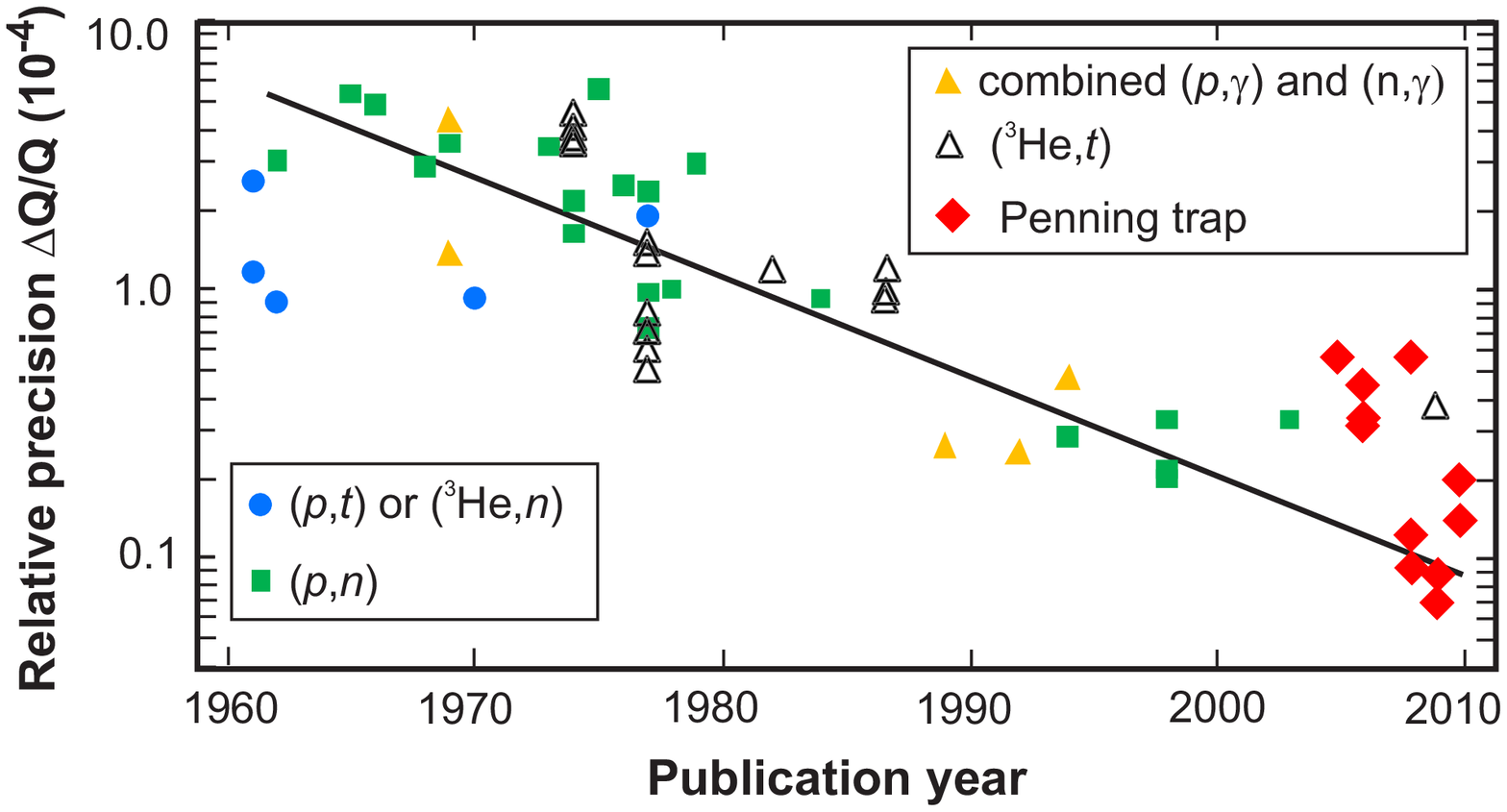}
 \caption{\label{fig:history}The relative precision, $\Delta Q/Q$, for $Q_{EC}$-value measurements of superallowed transitions is plotted
against their publication date, where $Q$ is the measured $Q_{EC}$ value and $\Delta Q$ is its quoted uncertainty.  The data encompass the
superallowed transitions from \nuc{10}{C}, \nuc{14}{O}, \nucm{26}{Al}, \nuc{34}{Cl}, \nucm{38}{K}, \nuc{42}{Sc}, \nuc{46}{V}, \nuc{50}{Mn}
and \nuc{54}{Co}, and are taken from a series of survey articles \cite{Ha09,Ha75,Ha90,Ha05} plus two more-recent publications \cite{Er09,Er11}.
Each point is identified by the experimental method used in the corresponding measurement. The line simply illustrates the decreasing trend.
(Adapted from ref.\,\cite{Er12}.)}
\end{figure}

The relative precision, $\Delta Q/Q$, obtained for various measurements and techniques is plotted in
Fig.\,\ref{fig:history} as a function of publication date, starting in 1960.  Evidently the relative precision improved steadily over the years,
from $\sim$5$\times$$10^{-4}$ in the early 1960s to $\sim$3$\times$$10^{-5}$ by 1990, but at that point a limit seemed to have been reached: for
the next fifteen years there were no further improvements in $Q_{EC}$-value precision, not even with the appearance of the first few Penning-trap
measurements.  However, the figure shows that in the space of only a few years after their first contribution, Penning traps had improved the relative
precision for measured $Q_{EC}$ values by a factor of 5, to as low as $\sim$7$\times$$10^{-6}$.  To put this last number into perspective, it
applies to the decay of \nucm{38}{K} and corresponds to an uncertainty of $\pm$40 eV on a total $Q_{EC}$ value of 6044.22 keV \cite{Er09}.

Although Penning traps are now outstripping nuclear reactions in the precision of their $Q_{EC}$-value results, for each transition it is
the average of all measurements with uncertainties within a factor of ten of the best measurement that is used in the determination of $V_{ud}$.
Many reaction measurements therefore still make significant contributions to the world averages for $Q_{EC}$ values.  We will describe several
examples of important reaction measurements before doing the same for Penning-trap measurements.

\subsubsection{\label{pnreact} The ($p$,$n$) reaction}

It can easily be seen in Fig.\,\ref{fig:history} that before 2005 the dominant choice for making $Q_{EC}$-value measurements of $0^+$$\rightarrow 0^+$
transitions was the ($p$,$n$) reaction.  There are a total of 21 such cases documented in the figure, with all but one -- the first -- being
threshold measurements.  Between 1962 and 1976, all but two of the measurements were made by Freeman and her collaborators using the 12 MeV Tandem
accelerator at Harwell in England (e.g.~\cite{Fr66,Sq76}). From 1977 on, all but two were made by Barker {\it et al.}~using the folded Tandem accelerator
AURA2 at the University of Auckland in New Zealand (e.g.~\cite{Ba77,To03}).

The Freeman measurements and the early Barker measurements all used the same general approach: For a superallowed decay P$\rightarrow$D they bombarded
a thin target of the daughter material D with a proton beam for a well-determined time, and then interrupted the beam while they determined the amount
of the parent P that had been produced by recording the characteristic activity of P, usually its emitted positrons.  This beam on-off cycle was repeated
until sufficient statistics had been accumulated.  Then the process was repeated at a succession of different beam energies until a threshold curve for the
production of P had been obtained.  They calibrated the proton beam energy near threshold by scattering the beam at 90$^\circ$ from a thin gold foil into
a broad-range magnetic spectrograph, where it was compared with a known $\alpha$-particle group from the decay of a standard source such as \nuc{212}{Po}.
This, of course, meant that the accuracy of their threshold energy -- and the resultant $Q_{EC}$ value -- relied upon the accuracy of the $\alpha$-particle
energy as it was known at the time but, by clearly indicating the $\alpha$ energy they used, they ensured that their result could be upgraded in future
whenever that calibration energy was improved.

Barker's group refined this technique by passing the primary proton beam through an Enge split-pole magnetic spectrograph, which was set at 0$^\circ$.
The beam path and width were constrained by a set of slots and apertures, resulting in an energy profile with FWHM (full width half maximum) of 50-100
ppm (parts per million) at the focal plane of the spectrograph \cite{To03}.  It was this prepared beam that bombarded the target, which was located 50 cm
beyond the focal plane.  To determine the energy of the proton beam, the magnetic rigidity of the spectrograph orbit was calibrated by leaving the field
strength unchanged and passing a monoenergetic heavy-ion beam of cesium or potassium around the same constrained orbit.  That heavy-ion beam was produced
by surface ionization and then accelerated through a potential $V$, which was adjusted until the ions were observed just upstream from the target.  Finally,
the optimized voltage $V$ was compared with a 1-volt standard via two successive stages of resistive division.  The threshold determined from a yield curve
was thus based firmly on a primary calibration standard -- the volt -- independent of any secondary reaction $Q$ values.  However it did require corrections
for finite energy spread of the beam, for non-uniform proton energy loss and for atomic excitation.  It was also subject to near-threshold resonances, which,
if present, could misguide the analysis \cite{Br94,Er12}.  Nevertheless, at its apotheosis, this technique achieved a precision of $\sim$2$\times$$10^{-5}$
on several measured $Q_{EC}$ values.

\subsubsection{\label{pgreact} Combined ($p$,$\gamma$) and ($n$,$\gamma$) reactions}

A second reaction-based approach that ultimately led to very precise results was to measure both ($p$,$\gamma$) and ($n$,$\gamma$) reactions on a common
target, which had been chosen so that the ($p$,$\gamma$) reaction would produce the parent of a $0^+$$\rightarrow 0^+$ $\beta$ decay, and the ($n$,$\gamma$)
reaction would produce the daughter: for example, \nuc{25}{Mg}($p$,$\gamma$)\nuc{26m}{Al} and \nuc{25}{Mg}($n$,$\gamma$)\nuc{26}{Mg} \cite{Ki91}.  In such
cases, the reaction $Q$ values yield $Q_{EC}$ through the relation
\be
Q_{EC} = Q_{n\gamma} - Q_{p\gamma} - 782.347~{\rm keV,}
\label{Qpgpn}
\ee   
which is independent of the mass of the target nucleus.

Continuing with the same example \cite{Ki91}, the \nuc{25}{Mg}($n$,$\gamma$)\nuc{26}{Mg} reaction was studied with thermal neutrons from the Los Alamos
Omega West reactor.  Gamma rays were detected with a Compton-suppressed Ge(Li) detector, and their energies precisely determined from a calibration based
on well known lines observed in the \nuc{1}{H}($n$,$\gamma$), \nuc{12}{C}($n$,$\gamma$) and \nuc{14}{N}($n$,$\gamma$) reactions.  The value of $Q_{n\gamma}$
was then obtained from the average summed energy of a number of $\gamma$-ray cascades de-exciting the capture state in \nuc{26}{Mg}.  For the
\nuc{25}{Mg}($p$,$\gamma$)\nuc{26m}{Al} reaction a dual Mg-Al target (upper half Mg, lower half Al) was employed, over which a proton beam from the Utrecht
3 MV Van der Graaff accelerator was wobbled up and down.  The proton energies at four \nuc{25}{Mg}($p$,$\gamma$)\nuc{26m}{Al} resonances were compared with
four accurately known resonances in the \nuc{27}{Al}($p$,$\gamma$)\nuc{28}{Si} reaction.  The proton beam was scanned in 200-eV steps over each \nuc{25}{Mg}
resonance and its nearest neighbour \nuc{27}{Al} resonance, which in all four cases was only a few keV away.  Thus each of the proton energy differences
could be determined precisely and an average $Q_{p\gamma}$-value obtained.  Obviously this method for determining $Q_{EC}$ depended on secondary calibration
standards but it was less dependent on experimental corrections than was the ($p$,$n$) reaction described in Sec.~\ref{pnreact}.  At its best, it achieved a
relative precision of 3\,-\,5$\times$$10^{-5}$.

\subsubsection{\label{3het} Relative ($^3\mathrm{He},t$) reactions}

Like ($p$,$n$), a ($^3$He,$t$) reaction acting on the daughter of a superallowed $0^+$$\rightarrow 0^+$ decay produces the parent of the decay, so the reaction
$Q$ value is directly related to $Q_{EC}$.  In contrast with ($p$,$n$) though, the energy of the outgoing particle -- a tritium ion -- can be measured
conveniently.  Even so, to determine the $Q$ value, the energies of both the tritons and the $^3$He projectiles must still be calibrated at high precision
against an established energy standard, not an easy task at the energies involved.  Koslowsky, Hardy and collaborators at Chalk River dealt with this
difficulty by developing a novel system that allowed them to measure the $Q$-{\it value differences} between two ($^3$He,$t$) reactions produced concurrently from a
target containing two components, each the daughter of a superallowed $\beta$ emitter \cite{Ko87}.

\begin{figure}[t]
\hspace{2.3cm}
 \includegraphics[width=11cm]{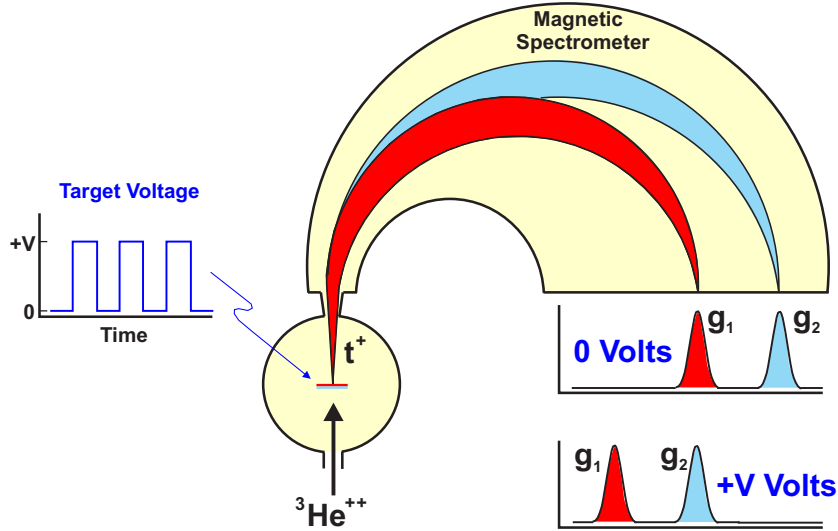}
 \caption{\label{fig:doublet}Schematic diagram illustrating the principle of biasing a two-component target in order to peak-match the reaction-product group $g_1$
from one component with group $g_2$ from the other component. (Adapted from ref.\,\cite{Ko87}.)}
\end{figure} 

Their method is illustrated in Fig.\,\ref{fig:doublet}.  A two-component target was bombarded by a doubly ionized $^3$He beam of 20\,-\,30 MeV from an MP Tandem
accelerator.  The ejected tritons were analyzed at 0$^\circ$ by a high-resolution Q3D magnetic spectrometer, which transformed their energy spectrum into a
distribution in position along its focal plane.  The target assembly was constructed so that it could be intermittently biased at $+V$ volts relative to its grounded
surroundings, with $V$ being adjustable up to 150 kV.  With the target at voltage $+V$, the $^3$He beam, being doubly ionized, was retarded by 2$V$ eV, while the
singly-charged tritons were re-accelerated by only $V$ eV.  As a result, the net effect of the imposed voltage was to reduce the triton energy by $V$ eV relative
to its value with no voltage on the target, thus shifting its position on the focal plane.  The target voltage could then be adjusted until the shifted position of
triton group $g_2$ exactly coincided with the unshifted position of $g_1$ (see Fig.\,\ref{fig:doublet}).  That adjusted voltage, which was related by resistive
division to the standard volt, corresponded to the energy difference between groups $g_1$ and $g_2$.  Furthermore, since the shifted tritons ($g_2$) followed the
same path as the unshifted ones ($g_1$), the result was independent of that path.  With $g_1$ chosen to be a triton group from one target component and $g_2$ being
from the other, the difference between reaction $Q$ values could be precisely determined with reference to the standard volt.

Of course, the superallowed $Q_{EC}$-value differences themselves were not 150 keV or less.  In practice, for each doublet, the experimental team determined the
$Q$-value difference for the population of an excited state in each of the two $\beta$-decay parents, these states being chosen so that their $Q$-value difference
was indeed within 150 keV.  Their excitation energies were either known or determined separately via $\gamma$-ray spectroscopy, so the measured reaction $Q$-value
differences could be related quite precisely to the superallowed $Q_{EC}$-value differences.  Four pairs of superallowed decay energies were studied in this way
\cite{Ko87}.  Being differences, the results do not appear in Fig.\,\ref{fig:history}, but their precision was comparable to the best reaction results that do
appear there, and they continue to figure prominently in the world averages for $Q_{EC}$ values \cite{Ha09}.

\subsubsection{\label{Penning} Penning-trap mass measurements}

The $Q_{EC}$ value for $\beta$ decay is simply the atomic mass difference between the parent state and its daughter: It could be derived from those masses if they were
known precisely enough.  So far, though, we have just described methods for measuring the difference directly since, until recently, this approach yielded the only
precise results.  As explanation, consider the superallowed $0^+$$\rightarrow 0^+$ decay, $^{26m}$Al$\rightarrow$$^{26}$Mg.  The masses of the nuclear states in
this case are $\sim$$2.4\times10^{10}$ eV and the difference between them ($Q_{EC}$) is $\sim$$4.2\times10^6$ eV.  Reaction measurements yield $Q_{EC}$-values with
relative precision of $\sim$$3\times10^{-5}$ (see Fig.\,\ref{fig:history}), which is $\sim$120 eV in this case.  To achieve the same precision with a pair of mass
measurements would require them each to have a relative precision of $\sim$$4\times10^{-9}$.  This was beyond the capability of conventional mass spectrometry,
which in any case was limited to effectively stable nuclei.

The balance has shifted with the appearance of Penning traps coupled on-line to an accelerator.  The Penning trap itself can confine charged particles to a small
volume by means of static magnetic and electric fields, the former being homogeneous and the latter quadrupolar.  The trapped ions exhibit three eigenmotions:
one along the axis of the magnetic field, and the other two in the radial plane perpendicular to that axis.  By combining the frequencies of these three
eigenmotions, one can obtain the cyclotron frequency, $\nu_c$, of the trapped ions and, from it, the mass of the ion itself since
\be
\nu_c = \frac{qB}{2\pi m},
\label{cycf}
\ee   
where $q$ and $m$ are the charge and mass of the ion, and $B$ is the magnetic field.  For stable nuclei, the frequency can be determined via external circuitry,
without the trapped ions being released \cite{Br86}.  A relative precision of 2-3$\times10^{-12}$ has been achieved this way for the stable nuclei, $^{32}$S and
$^{31}$P, for example \cite{Re08}.

\begin{figure}[t]
\hspace{2.3cm}
 \includegraphics[width=10cm]{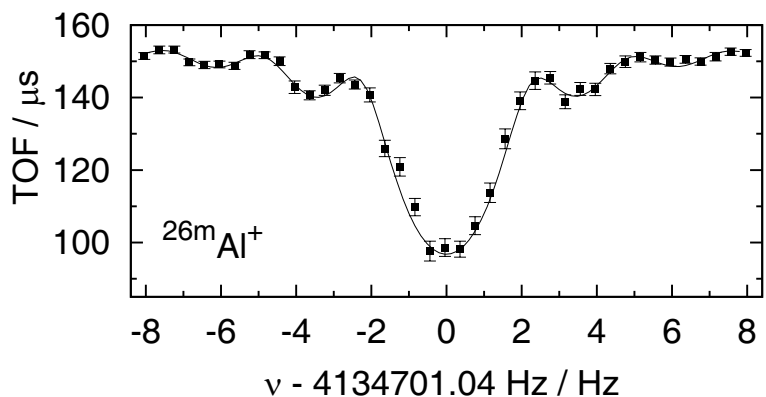}
 \caption{\label{fig:Penning}Time of flight (TOF) resonance measured with an on-line Penning trap for the superallowed emitter $^{26m}$Al, which has a half-life of
6.3 s.  The solid curve is a fitted function. (Adapted from ref.\,\cite{Er06}.)}
\end{figure} 

However, the parent nuclei of the superallowed decays -- and sometimes the daughters as well -- have short half-lives, from a few seconds down to a few
tens of milliseconds.  This requires a number of additional steps in the experimental procedures.  First the ions of interest must be produced by an accelerator;
next they are cooled, bunched and if necessary purified; and then they are injected into the Penning trap.  Because of the short half-life of the ions, this cycle
is repeated continuously with fresh ion bunches being delivered every few seconds.  In each cycle, once the ions are trapped, the cyclotron frequency in the
trap is probed with an applied radiofrequency electric field, after which the ions are released and their time of flight measured to a microchannel-plate
detector located outside the high-field region.  As the applied frequency is scanned through the cyclotron frequency, the ions' time-of-flight passes through a
distinct minimum, as is shown in Fig.\,\ref{fig:Penning}.  To this basic technique numerous refinements have been applied, the most significant of which is to excite the
ion motion with Ramsey's method of time-separated oscillatory fields \cite{Ra90}.  

On-line Penning trap measurements for superallowed $\beta$ decay have been, and are being performed at four different facilities: ISOLTRAP ({\it e.g.}~Ref.\,\cite{Mu04}),
CPT ({\it e.g.}~Ref.\,\cite{Sa05}), LEBIT ({\it e.g.}~Ref.\,\cite{Ri07}) and JYFLTRAP ({\it e.g.}~Refs.~\cite{Er09,Er11}).  An early high-impact measurement made with
the CPT Penning trap was of the $Q_{EC}$ value for the decay of $^{46}$V \cite{Sa05}.  The result, which was determined with an uncertainty of $\pm$400 eV, differed by
2.5 keV from a long-trusted 1977 ($^3$He,$t$) reaction measurement that had claimed a similar precision.  The latter measurement was based on a
``precision time-of-flight measuring system" with the Q3D spectrograph at the Munich Tandem Laboratory; it was one of seven superallowed $Q_{EC}$ values that appeared in
a single publication \cite{Vo77} and had stood unchallenged for nearly 30 years.  However, the $^{46}$V discrepancy was soon confirmed by a second Penning trap, JYFLTRAP, which
also identified similar disagreements with the Munich measurements for three other cases, $^{42}$Sc, $^{50}$Mn and $^{54}$Co \cite{Er06,Er08}.  When a repeat ($^3$He,$t$)
measurement \cite{Fa09}, made at Munich in 2009 with much of the same equipment, agreed with the Penning trap results, it was decided to eliminate Ref.\,\cite{Vo77} from
surveyed world data \cite{Ha09}.  Fortunately, this was the only significant disagreement found between Penning-trap results and those from earlier reaction measurements.
It has since been demonstrated that one can safely combine the results of both types of measurement without including any additional systematic uncertainties \cite{Er12}.

To date, the most precise Penning-trap measurements of superallowed $Q_{EC}$ values have been done by Eronen and collaborators 
with the JYFLTRAP trap at the University of Jyvaskyla, where it is coupled to a cyclotron through their Ion Guide Isotope Separator On Line (IGISOL) \cite{Er12}.
In addition to all the refinements to achieve high trap precision, this facility has an added advantage: It can produce both parent and daughter nuclei with
the same beam.  For the case illustrated in Fig.\,\ref{fig:Penning}, the superallowed emitter $^{26m}$Al was produced by the ($p$,$n$) reaction at 15\,MeV on a target
of $^{26}$Mg.  Ions of $^{26}$Mg, which is the $\beta$-decay daughter, were also released by elastic scattering of the proton beam.  The $Q_{EC}$ value is
then given by
\be
Q_\mathrm{EC} = M_p - M_d =  \left ( \frac{\nu_{c,\mathrm{d}}}{\nu_{c,\mathrm{p}}} -1 \right ) \left (M_d - m_e \right) + \Delta_{p,d},
\label{qec}
\ee
where $M_p$ and $M_d$ are the parent and daughter masses, and $\nu_{c,\mathrm{p}}$ and $\nu_{c,\mathrm{d}}$ are their respective measured cyclotron frequencies;
$m_e$ is the electron rest mass; and the term $\Delta_{p,d}$ arises from atomic-electron binding-energy differences between the parent and daughter, known to
sub-eV accuracy.  Since the term $(\nu_{c,\mathrm{d}}/\nu_{c,\mathrm{p}} - 1)$ in (\ref{qec}) is always $\ll$$10^{-3}$ for the superallowed parent-daughter
pairs, $M_d$ needs only to be known to few-keV precision in order for its uncertainty to have a negligible impact on the $Q_{EC}$-value precision.

Recent $Q_{EC}$ measurements with JYFLTRAP interleave parent and daughter frequency measurements by switching automatically back and forth between parent and
daughter ions after each complete frequency scan, typically every minute or so.  This effectively eliminates any systematic differences that might occur from
drifts in the magnetic field for example.  As a result, a relative precision of $7\times10^{-6}$ has been achieved in several cases, including the $Q_{EC}$ value
for the superallowed branch from $^{38m}$K, which was determined to be 6044.223(41)\,keV \cite{Er09}.

\subsection{\label{t} Half-lives}

Precise half-life measurements are deceptively difficult.  Problems such as impurity activities, rate-dependent thresholds, dead-time and pile-up effects, as well
as statistically flawed analyses, offer no obvious signals of their magnitude, or even of their presence.  It is not surprising that many half-life measurements of
superallowed emitters have had to be rejected from surveys of world data (see table VII in Ref.~\cite{Ha09}).

\begin{figure}[t]
\hspace{2.3cm}
 \includegraphics[width=10cm]{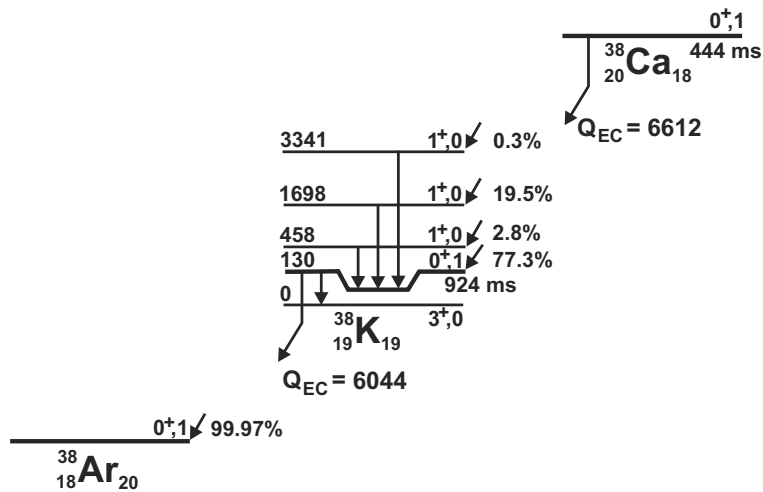}
 \caption{\label{fig:38decays}Partial decay schemes of $^{38}$Ca and $^{38m}$K.}
\end{figure}

The superallowed $0^+$$\rightarrow 0^+$ transitions we are considering here take place between $T$\,=\,1 analogue states.  As shown in Fig.\,\ref{fig:chart}, the parents
are of two types: either odd-$Z$-odd-$N$ nuclei with $T_Z$\,=\,0, or even-even ones with $T_Z$\,=\,$-1$.  The two types exhibit quite different decay patterns, as
is shown in Fig.\,\ref{fig:38decays}, where $^{38m}$K$\rightarrow ^{38}$Ar is an example of the first type and $^{38}$Ca$\rightarrow ^{38}$K is an example of the second.  Not
surprisingly, most of the best-measured decays in the past have been of $T_Z$\,=\,0 nuclei, where the superallowed branch is overwhelmingly predominant.  For such
cases the only way to measure the half-life is to detect the emitted positrons (or possibly the 511-keV annihilation radiation).  However, for the decays of $T_Z$\,=\,$-1$
nuclei -- two of which are the classic cases of $^{10}$C and $^{14}$O, discussed in the Introduction -- the total $\beta$-decay strength is spread over a number of
branches and ample $\beta$-delayed $\gamma$ rays are produced.  In these cases half-life measurements based on $\gamma$-ray detection have been reported as well
as those in which only positrons were recorded.  We will briefly describe both measurement techniques with their advantages and disadvantages.

\subsubsection{\label{beta} Beta detection methods}

Direct detection of decay positrons can be accomplished with high efficiency, and the signals from the detector -- either a plastic scintillator or a gas counter -- 
can be processed safely at quite high rates.  Against these advantages must be balanced the disadvantage presented by the positrons' broad energy distribution, which
cannot in general be used to distinguish one decaying nuclide from another.  Without some external means of ensuring source purity, a decay measurement can easily be
invalidated by the presence of an undetected impurity.  Typically, before 1983, activities were produced by low-energy proton beams on enriched targets, a combination that
minimized contaminant activities but could not eliminate them entirely.  In a few cases, where impurities were identified, their contribution was corrected for ({\it 
e.g.}~Refs.\,\cite{Ry73a,Az75}) but, in most cases, purity was simply a fervent belief.   Since 1983 however, with rare exceptions all measurements have employed sources
produced via either an on-line isotope separator -- first at Chalk River ({\it e.g.}~Ref.\,\cite{Ko83}) and later at TRIUMF ({\it e.g.}~Ref.\,\cite{Fi11}) -- or the
magnetic recoil spectrometer at the Texas A\&M cyclotron ({\it e.g.}~Ref.\,\cite{Pa12}).  These devices eliminate or at least minimize impurities, with the recoil
spectrometer also being capable of identifying and quantifying any weak impurities that remain.

\begin{figure}[t]
\hspace{2.3cm}
 \includegraphics[width=12cm]{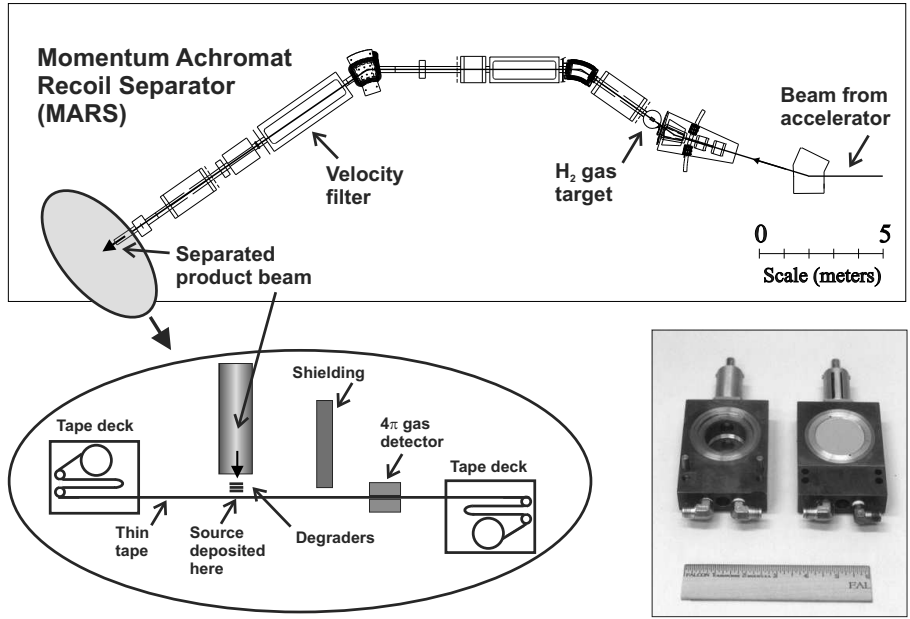}
 \caption{\label{fig:TAMUexp}Simplified experimental arrangement used, for example, in measuring the half-lives of \nuc{38}{Ca} \cite{Pa11} and \nuc{46}{V} \cite{Pa12}.
The two halves of the 4$\pi$ gas proportional counter are pictured at the bottom right.  In operation, each half is sealed with a Havar window (as shown on the right
half only) and the two halves are bolted together with the Havar windows facing one another.  There is a thin slot that remains so the tape can pass between the windows.}
\end{figure}

Although half-life measurements in the past have frequently used plastic scintillators to detect $\beta$ particles, in all but one of the thirteen cases of superallowed decays that
currently contribute to world data \cite{Ha09}, the most precise half-life measurements have all been made with 4$\pi$ gas proportional counters, all built from the
design first developed for this purpose by Koslowsky, Hagberg and Hardy  at Chalk River 30 years ago \cite{Ko83}.  It was modeled after the ``pill box" detectors long
used by radiation metrologists \cite{NC85}, chosen because they have low background and are nearly 100\% efficient for $\beta$ particles, while being insensitive to
$\gamma$ rays.  The detector consists of two separate gas cells machined from copper (as pictured in Fig.\,\ref{fig:TAMUexp}), each containing an anode of gold-plated tungsten
wire 13 $\mu$m in diameter, and each hermetically sealed by a Havar window 3.7\,cm in diameter and 1.5\,$\mu$m thick.  Methane at just above atmospheric pressure is
continuously flushed through both cells.  When assembled together, there is a 25-mm slot between the two cell windows, through which a thin tape can slide.  The assembled
detector is easily held in the palm of one hand.

Though the detector itself is small, the equipment required to deliver a clean source into the detector is not.  In the experimental configuration employed at Texas A\&M
(see top and left side of Fig.\,\ref{fig:TAMUexp}) the activity of interest is first produced by bombardment of a cooled hydrogen gas target.  Taking the superallowed
parent \nuc{46}{V} ($t_{1/2}$\,=\,423 ms) as an example \cite{Pa12}, the activity was produced from the \nuc{1}{H}(\nuc{47}{Ti},\,2$n$)\nuc{46}{V} reaction initiated by a
1.5-GeV beam of \nuc{47}{Ti} from the K500 cyclotron.  The fully stripped reaction products exiting the gas cell entered the Momentum Achromat Recoil Spectrometer (MARS)
\cite{Tr89}, where they were separated according to their charge-to-mass ratio $q/m$, with \nuc{46}{V} being selected by slits in the focal plane.  A position-sensitive
silicon detector was periodically inserted at the focal plane to identify and monitor any weak contaminants that were also passing through the slits.

The purified \nuc{46}{V} beam was extracted into air, degraded and implanted into the 76-$\mu$m-thick aluminized Mylar tape of a fast tape-transport system.  The combination
of $q/m$ selectivity in MARS and range sensitivity in the degraders led to collected samples of radioactive \nuc{46}{V}, in which the only interfering activity was
determined to contribute less than 0.012\% to the total.  After a sample had been collected for 0.5\,s, the beam was turned off and the tape moved the sample 90 cm into
the center of the shielded 4$\pi$ gas detector, where it stopped less than 200\,ms later.  Signals from the detector were then multiscaled for 10\,s, which is more than 20
half-lives of \nuc{46}{V}.  This cycle was repeated many thousands of times until the desired statistics had been reached.

In all measurements of this type, the counting electronics impose a well-defined non-extendable dead time.  This dead time as well as other measurement parameters, such
as detector bias and discriminator levels, are altered from time to time in order to test for possible systematic effects.  Careful analysis and fitting procedures are
applied and these procedures are checked with hypothetical data, computer-generated by Monte Carlo techniques to simulate closely the experimental counting conditions.

In recent years this system has also been used to measure the half-lives of $T_Z$\,=\,$-1$ superallowed parents such as \nuc{38}{Ca} (see Fig.\,\ref{fig:38decays}).  There
a complication arises: The $\beta$ decay of the $T_Z$\,=\,$-1$ parent feeds a second superallowed decay, from the $T_Z$\,=\,0 daughter to the $T_Z$\,=\,+1
granddaughter.  Of course, the positrons from both decays are recorded simultaneously in the $\beta$ detector.  The time-decay spectrum is therefore the sum of the decay
of the parent and the growth-and-decay of the daughter.  Typically, the parent's half-life is about a factor of two shorter than the daughter's, so the decay of the
daughter almost completely masks the decay of the parent ({\it e.g.}~see Refs.\,\cite{Pa11,Ia10}).  However the composite decay can still be used to determine
the parent's half-life if three conditions are met: 1) the half-life of the daughter is known precisely from an independent measurement; 2) the $T_Z$\,=\,$-1$ parent activity
deposited on the transport tape is pure; and 3) the rate at which it is deposited is known.  The first condition can easily be met by a measurement of the type already
described for \nuc{46}{V}; the second is assured by the combination of electromagnetic separation (MARS) and range selectivity; and the third was met by insertion of a
thin scintillator into the degrader stack (see Fig.\,\ref{fig:TAMUexp}), from which the number of ions were recorded as a function of time during sample collection.  

With these methods, half-lives with a relative precision of $\sim$0.03\% for $T_Z$\,=\,$-1$ parents ({\it e.g.}~\nuc{26}{Si} \cite{Ia10}) and $\sim$0.01\% for $T_Z$\,=\,0
parents ({\it e.g.}~\nuc{46}{V} \cite{Pa12}) have been obtained in the best cases.  A result of comparable precision was also obtained for \nucm{26}{Al} at TRIUMF \cite{Fi11},
where the same type of gas counter was used but with the sample produced by their isotope separator ISAC.  Very recently another measurement of the \nucm{26}{Al} half-life with
similar precision has been reported \cite{Ch13}, for which digital pulse-analysis was used to process signals from the 4$\pi$ gas detector instead of analogue electronics.
This meant that parameters like dead-time and discriminator level could be investigated, after the fact, on the saved pulse shapes, thus improving the efficiency of
data-taking.  The relative precision quoted for this measurement was also at the 0.01\% precision level, with some promise for future improvement.

\subsubsection{\label{gamma} Gamma detection methods}

For a half-life measurement, high-resolution $\gamma$-ray detection makes source purity a much less critical requirement, since analysis can focus on the photopeak of
a $\gamma$ ray that is characteristic of the activity of interest.  This is the method's principal advantage.  On the other side of the ledger must be placed the
relatively low efficiency of germanium detectors and the slow signals that are derived from them, with the consequently long time required to process those signals.
The latter introduces uncertainties in accounting for dead time and especially for pulse pile-up, which is of course rate dependent.  A further disadvantage is that
the method can only be applied to $T_Z$\,=\,$-1$ parents of superallowed decays since they are the only ones that produce $\beta$-delayed $\gamma$ rays of sufficient
intensity (see Fig.\,\ref{fig:38decays}).

A recent measurement at TRIUMF illustrates the inherent difficulties with this method. Using a separated beam from the ISAC facility, Laffoley {\it et al.}~\cite{La13}
measured the half-life of \nuc{14}{O}.  This nucleus $\beta$-decays to \nuc{14}{N} with a 99.4\% branch to the 2313-keV excited state, which then de-excites by emitting
a $\gamma$ ray to the ground state.  The \nuc{14}{O} half-life can then be measured through detecting either $\gamma$ rays or $\beta$ particles.  Laffoley {\it et al.}~did
both simultaneously.   They implanted \nuc{14}{O} from ISAC into thin aluminum at the center of the ``8$\pi$" $\gamma$-ray spectrometer, a spherically symmetric array of 20
HPGe detectors.  A fast plastic scintillating detector was placed immediately behind the implantation location to detect $\beta$ particles.  The $\gamma$-ray data (for
the 2313-keV transition) were carefully analyzed with well worked-out techniques \cite{Gr07} to account for pile-up and other time-dependent effects.  The $\beta$
signals were handled in a very similar way to that used with a gas proportional counter.  The results are revealing: The half-life obtained from $\gamma$ counting was
$70.632 \pm 0.086_{stat} \pm 0.037_{syst}$\,s, while from $\beta$ counting it was $70.610 \pm 0.020_{stat} \pm 0.023_{syst}$\,s.  Though the two measurements were made
simultaneously on the same collected sources, both the statistical and systematic uncertainties were larger when $\gamma$-rays were employed.  Happily though the
two results agreed with one another well within their quoted uncertainties.

\subsection{\label{R} Branching ratios}

Of the three experimental quantities -- $Q_{EC}$ values, half-lives and branching ratios -- needed to obtain an $ft$ value, the most difficult to measure precisely is
the branching-ratio.  Since the continuous energy distribution of emitted positrons leaves little opportunity to distinguish one transition from another, such
measurements must be based on detection of the $\beta$-delayed $\gamma$ rays emitted from levels populated in the $\beta$-decay daughter.  To make matters more
difficult, in most cases one of the $\beta$ transitions populates the ground state or an isomeric state, from which no $\gamma$-ray signal is forthcoming.  Thus it
is not enough to measure the relative branching to excited states.  What is needed is the absolute branching for each transition: {\it i.e.}~the fraction each accounts
for out of the total decays of the parent nucleus.

What are the competing transitions?  Since the parents of our decays of interest have spin-parity $0^+$, they can populate $1^+$ states in the daughter by allowed
Gamow-Teller decay, in addition to populating the analogue $0^+$ state by the superallowed (Fermi) branch.  Furthermore, weak Fermi branches are also possible to
excited $0^+$ states via charge-driven mixing with the analogue state.  The nuclear structure of the daughter nucleus and the energy available for the parent's $\beta$
decay together determine how many such states can be populated.

Here again the challenges are different for the decays of $T_Z$\,=\,0 superallowed parents compared to the $T_Z$\,=\,$-1$ cases.  The superallowed branch from each
$T_Z$\,=\,0 parent carries $>$99\% of the decay strength and populates the ground state of its daughter (see Fig.\,\ref{fig:38decays}).  To determine its exact
branching ratio, all that is required is to measure the weak competing branches, if any, with modest precision and subtract their total from 100\%.  Since the absolute
value of the uncertainty on the total of the weak branches becomes the uncertainty on the superallowed branch, a poor relative precision on the former becomes
a very good relative precision on the latter.  For example, the decay of \nuc{42}{Sc} includes a single competing Gamow-Teller $\beta$-decay branch to the 1.84-MeV state
in \nuc{42}{Ca}.  Its branching ratio, based on four separate measurements, is 0.0059(14)\% \cite{Ha09}, a result with $\pm$24\% relative precision.  The superallowed
branching ratio obtained from this result is 99.9941(14), which has a relative precision of $\pm$0.0014\%!  Obviously, high precision is not the issue with these
measurements.  Rather, the difficulty lies in even observing branches with relative intensities that are less than 100 parts per million of the total decay.

The decays of $T_Z$\,=\,$-1$ superallowed parents are quite different.  In general they are characterized by much stronger competing Gamow-Teller transitions (see
Fig.\,\ref{fig:38decays}) and, in a few cases, the superallowed branch is not even the strongest transition: In \nuc{10}{C} decay, for example, the superallowed branch
only accounts for 1.4646(19)\% of the total decay strength \cite{Ha09}.  In addition, with two exceptions -- \nuc{10}{C} and \nuc{22}{Mg} -- all the known cases include
decay branches that do not produce a subsequent $\gamma$ ray.  Consequently, absolute branching ratios must be determined, but without nearly the precision improvement
factor just described for decays like that of \nuc{42}{Sc}.  These ratios must therefore be directly determined with $\sim$0.1\% precision.   Only very recently has it
become possible to do so. 

\subsubsection{\label{TZ0} $T_Z$\,=\,0 parent decays}

Because the daughter of a $T_Z$\,=\,0 superallowed parent is an even-$Z$-even-$N$ nucleus, its excited $1^+$ and $0^+$ states are at a relatively high excitation energy above
the $0^+$ ground state, which is strongly populated by the superallowed transition.  For the lightest nuclei, with $A$$\leq$\,38, the $\beta$-decay energy window is such
that none of these excited states is populated, the limit of observation being at $\sim$10 parts per million.  However, as $A$ increases, the $\beta$-decay energy of the
parent grows and the level density in the daughter increases; so competing branches become greater in number and in strength.  For \nuc{42}{Sc}, the lightest emitter for
which a non-superallowed $\beta$ branch has been observed, that branching ratio is 0.0059\%; while for \nuc{74}{Rb}, the heaviest well-measured case, there are a number of
competing branches, which total to a 0.50\% branching ratio.

There have been relatively few measurements of these weak non-superallowed branching ratios, since parts-per-million sensitivity is not easy to achieve.  One example is
the work of Hagberg and collaborators \cite{Ha94} at Chalk River, who investigated four emitters, \nucm{38}{K}, \nuc{46}{V}, \nuc{50}{Mn} and \nuc{54}{Co}.  The first was
produced by an ($\alpha$,$n$) reaction, the other three by ($p$,$n$) reactions, in all cases on isotopically enriched targets.  A helium-jet gas-transfer system was used
to convey each activity to a low-background counting location where the activity-loaded NaCl aerosol clusters in the helium were deposited onto the aluminized Mylar tape of a
fast tape-transfer system.  After a short collection period, typically $\sim$0.5\,s, a paddle was inserted between the helium-jet nozzle and tape, and the tape moved the
sample in sequence to two different detector stations, stopping at each; then the cycle was repeated until adequate statistics had been acquired.  To achieve the required
sensitivity, MBq-level sources were required for each cycle.

The first detector station consisted of two thin plastic scintillators located on either side of the tape, with an HPGe detector in close geometry behind one of them.  The
latter was passively shielded against the high flux of energetic positrons from the dominant ground-state branch.  Gamma-ray signals from the HPGe detector were only recorded
if they were in coincidence with $\beta$ signals from the opposite-side scintillator and in anti-coincidence with those from the same-side scintillator.  This singled out
true $\beta$-delayed $\gamma$ rays while eliminating bremsstrahlung radiation in the HPGe detector caused by $\beta$ particles backscattering from the opposite-side
plastic.

\begin{figure}[t]
\hspace{2.3cm}
 \includegraphics[width=12cm]{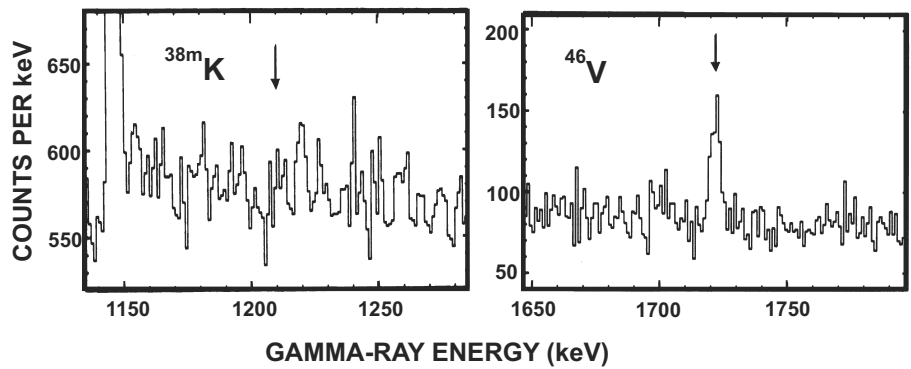}
 \caption{\label{fig:gspect}Portions of gated $\gamma$-ray spectra obtained following $\beta$ decays of \nucm{38}{K} and \nuc{46}{V} \cite{Ha94}.  The position of the possible
$0^+_1$$\rightarrow$\,$2^+$ $\gamma$ ray is indicated with an arrow in both cases. The strong peak in the \nucm{38}{K} spectrum is the double-escape peak from the 
2168-keV $\gamma$ ray from the $\beta$ decay of the \nuc{38}{K} ground state.  (Adapted from ref.\,\cite{Ha94}.)}
\end{figure}

At the second detector station the tape stopped in the center of a 4$\pi$ gas proportional counter (see Sec.\,\ref{beta}) with nearly 100\% efficiency for $\beta$ particles.
The multiscaled data from this detector were used to determine the strength of the source in each cycle.  With the total strength known, the branching ratio corresponding to
any $\gamma$-ray peak observed in the HPGe spectrum could be obtained.  The HPGe detector efficiency was calibrated with standard sources and there were also significant
dead-time and other corrections to be applied so only $\sim$10\% relative precision could be quoted on the result but, because the transitions were so weak in the first place,
that was more than sufficient.  Portions of the $\gamma$-ray spectra they recorded from \nucm{38}{K} and \nuc{46}{V} are shown in Fig.\,\ref{fig:gspect}.  The arrows indicate
where $\gamma$ rays from known excited $0^+$ states would appear if those states were populated by $\beta$ decay.  An upper limit of 19 ppm was obtained
for this possible non-superallowed branch from \nucm{38}{K}, and a value of 39(4) ppm was derived from the clearly observed peak in the case of \nuc{46}{V}.  Results were
also obtained for \nuc{50}{Mn} and \nuc{54}{Co}.

The situation becomes much more complex for $T_Z$\,=\,0 parents with $A$$\geq$\,62.  This is well illustrated by the work of Finlay {\it et al.}~\cite{Fi08} at TRIUMF, who
studied the $\beta$ decay of \nuc{62}{Ga}.  They identified 30 $\beta$-coincident $\gamma$ rays, which they attributed to non-superallowed $\beta$ transitions from \nuc{62}{Ga}
to 10 excited $0^+$ or $1^+$ states in its daughter \nuc{62}{Zn}.  To obtain this result they deposited \nuc{62}{Ga} ions from the ISAC separator onto aluminized Mylar tape at
the mutual centers of an array of 20 thin plastic scintillators with $\sim$80\% efficiency for $\beta$ detection, and the ``8$\pi$" $\gamma$-ray spectrometer, an array of 20 HPGe
detectors operated in Compton-suppressed mode.  Although data were recorded continuously, the beam was cycled on and off, with an implantation period lasting 30\,s sandwiched
between two shorter periods: one before for background counting, and one after for decay counting.  At the end of each cycle, a tape-transport system moved the collected sample
to a shielded location, leaving a fresh portion of tape for the next cycle.  More than 5000 cycles were recorded in all.

While this thorough experiment might seem to have quantified all possible decay branches, actually it cannot.  For these higher $A$ values, hundreds or even thousands of $1^+$
states in the daughter can become accessible to $\beta^+$/EC decay.  Although most of these transitions are undoubtedly very weak, their aggregate can be quite significant at the
level of precision required in these measurements.  This is the ``Pandemonium" effect, which was first described in 1977 in a more general context \cite{Ha77} and, more recently,
has been applied specifically to superallowed $\beta$ decay \cite{Ha02}.  In the case of \nuc{62}{Ga} decay, shell-model calculations predict that over a hundred $1^+$ states
in \nuc{62}{Zn} can be populated by $\beta$ decay \cite{Fi08}.  With only 10 identified, Finlay {\it et al.}~had to make a correction to their result to account for transitions
that they could not observe individually.  Their approach hinged on two low-lying $2^+$ states in \nuc{62}{Zn}, which cannot be populated by allowed $\beta$ decay yet were seen
to have more $\gamma$-ray intensity de-populating them than feeding them.  The missing feeding could only be attributed to the presence of a large number of $\gamma$-ray
transitions, each too weak to observe, from excited states fed by correspondingly weak Gamow-Teller $\beta$ transitions.  After a further adjustment from a comparison with theory,
they concluded that their detailed spectroscopy had only identified $\sim$94\% of the non-superallowed $\beta$ intensity, and they corrected their result accordingly, arriving at
a final superallowed branching ratio of 99.858(8)\%.

\subsubsection{\label{TZ-1} $T_Z$\,=\,$-$1 parent decays} 

\begin{figure}[t]
\hspace{2.3cm}
 \includegraphics[width=8cm]{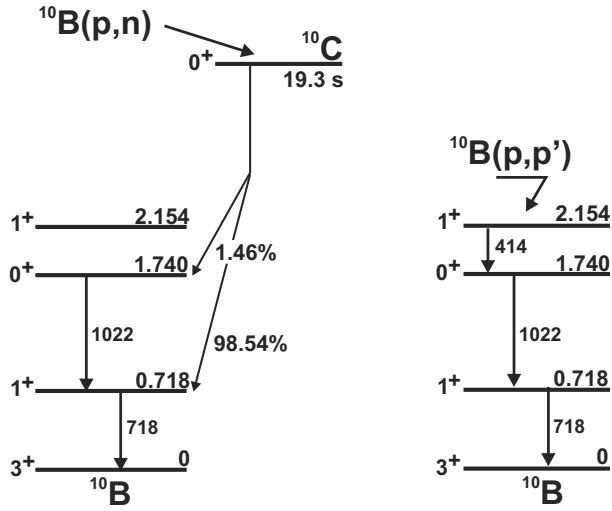}
 \caption{\label{fig:10C} Left: Decay scheme of \nuc{10}{C}, which is produced by the \nuc{10}{B}($p$,$n$) reaction.  Right: Main de-excitation route for the 2.154-MeV level in
\nuc{10}{B} populated by inelastic proton scattering on \nuc{10}{B}.  The level energies are given in MeV; the $\gamma$-ray energies are in keV.}
\end{figure}  

As described in the Introduction, the first $T_Z$\,=\,$-1$ parent decays to be studied were those of \nuc{10}{C} and \nuc{14}{O}.  Each has its own unique problem that even today
limits the precision with which its branching ratio can be measured.  The decay scheme of \nuc{10}{C} is shown on the left side of Fig.\,\ref{fig:10C}.  Since $\beta$ decay to
the $3^+$ ground state is second forbidden and decay to the $1^+$ state at 2.154 MeV is energetically disfavoured, the superallowed branching ratio is simply given by the ratio
of the number of $\gamma$ rays emitted at 1022 keV relative to the number at 718 keV.  There are two problems with this: the superallowed branch is weak and the energy of its
characteristic $\gamma$ ray is exactly twice that of the 511-keV annihilation radiation, which appears in abundance from the decay positrons.  Precision requires high statistics
together with confidence that the pile up of annihilation radiation has not contaminated the peak of interest.  These two conditions tend to work against one another.
  
Savard {\it et al.}~at Chalk River \cite{Sa95} dealt with these conflicting demands by using the array of 20 HPGe detectors that constituted the 8$\pi$ spectrometer.  This yielded
a twentyfold reduction in 511-511 pile-up compared with a single detector at the same total counting rate.  The experiment itself comprised two interleaved measurements.  One
was a repeated cycle in which the \nuc{10}{C} was first produced by the ($p$,$n$) reaction on a \nuc{10}{B} target mounted at the center of the spectrometer; then the beam was
interrupted and the $\beta$-delayed $\gamma$ rays from the decay were observed in singles mode.  The second measurement was performed in beam with $\gamma$-$\gamma$ coincidences
recorded from the deexcitation of the 2.154-MeV level in \nuc{10}{B}, which was populated by the ($p$,$p^{\prime}$) reaction (see right side of Fig.\,\ref{fig:10C}).  The ratio of
the number of $\gamma_{414}$-$\gamma_{718}$ coincidences to $\gamma_{414}$-$\gamma_{1022}$ coincidences in the second measurement determines the relative counting efficiencies for
the 718- and 1022-keV $\gamma$ rays, which can then be used to determine the relative $\gamma$-ray intensities in the first measurement.  In this way the superallowed branching
ratio was determined to be 1.4625(25)\%.  Subsequently, a similar measurement was made with the Gammasphere detector array \cite{Fu99}, which gave a slightly less precise, but
nevertheless consistent result.

The decay of \nuc{14}{O} has an even more challenging feature.  The superallowed branch carries more than 99\% of the decay strength and populates the analogue state at 2.31 MeV in
\nuc{14}{N}.  However, its strongest competition comes from a 0.6\% Gamow-Teller branch to the ground state, which emits no subsequent $\gamma$ ray to signal its appearance.
The only way to determine the precise strength of this ground state $\beta$-decay branch is to measure the energy spectrum of all the emitted positrons and tease out the contribution
of the ground state branch.  The last time this measurement was made was in 1966 \cite{Si66} (though the analysis was updated more recently \cite{To05}).  This measurement begs to be
repeated.

\begin{figure}[t]
\hspace{2.3cm}
 \includegraphics[width=12cm]{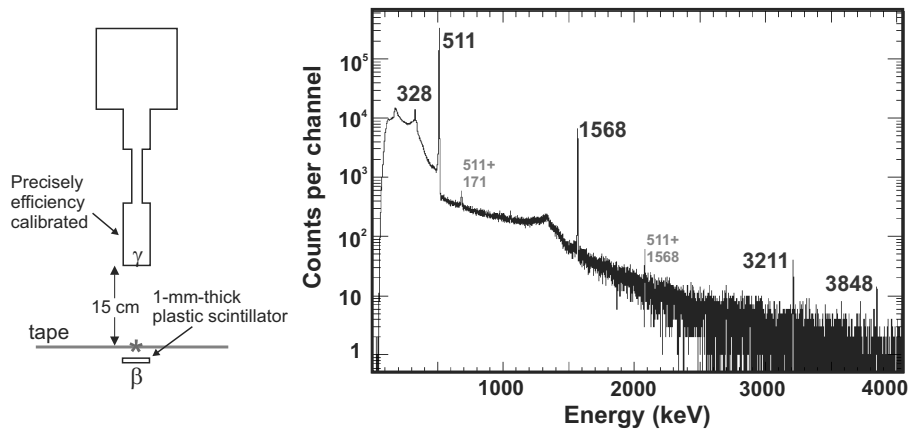}
 \caption{\label{fig:BR} Left: Arrangement of the $\beta$ and $\gamma$-ray detectors, between which the source is placed by the tape transport system illustrated in
Fig.\,\ref{fig:TAMUexp}.  Right: Spectrum of $\beta$-coincident $\gamma$ rays recorded from 61,000 cycles, each having a 1.54-s counting period \cite{Pa14}.}
\end{figure}

There are, of course, other $T_Z$\,=\,$-1$ superallowed parents with $A$$\geq$$18$ but, until very recently, their branching ratios have defied $\pm$0.1\% measurements since in
all cases but one, \nuc{22}{Mg}, not every $\beta$-decay branch feeds excited states that subsequently emit $\gamma$-rays.  With the ground state -- or a low-lying isomeric state --
also populated, the only way to arrive at correct branching ratios is to measure the intensity of the $\gamma$-ray peaks relative to the total number of decays of the parent
nucleus.  A method to achieve this has now been developed at Texas A\&M University by Hardy, Iacob and Park \cite{Pa14} using the same source-production and delivery system as
illustrated in Fig,\,\ref{fig:TAMUexp}.  For these measurements, though, the gas proportional detector shown in that figure is replaced at the counting station by a thin
plastic scintillator on one side of the tape and an HPGe detector on the other side, as shown on the left side of Fig.\,\ref{fig:BR}.  They record $\beta$ singles and
$\beta$-$\gamma$ coincidences.

If the $\gamma$ ray de-exciting state $i$ in the daughter is denoted by $\gamma_i$, then the $\beta$-branching ratio, $R_i$ for the $\beta$ transition populating that state
can be written:
\be
R_i = \frac{N_{\beta\gamma_i}}{N_{\beta}~\epsilon_{\gamma_i}} k,
\ee
where $N_{\beta\gamma_i}$ is the total number of $\beta$-$\gamma$ coincidences measured in the $\gamma_i$ peak, $N_{\beta}$ is the total number of $\beta$ singles,
$\epsilon_{\gamma_i}$ is the HPGe efficiency for detecting $\gamma_i$, and $k$ is a small correction factor ({\it i.e.}~$k$$\sim$1) that accounts for dead time and
pile-up, coincident summing, and the small changes in $\beta$-detector efficiency for the different energy transitions participating in the decay.  The equation highlights
the importance of having a pure sample -- so that $N_{\beta}$ can be relied upon -- as well as having a precise absolute efficiency calibration for the $\gamma$-ray detector,
and a reasonable knowledge of relative efficiencies in the beta detector.  The key ingredient that the Texas A\&M team has painstakingly developed is an HPGe detector
whose absolute efficiency has been accurately determined (to $\pm$0.2\% for 50-1400 keV $\gamma$ rays and to $\pm$0.4\% up to 3500 keV) from source measurements and Monte
Carlo calculations \cite{He03, He04}.

This method yields the branching ratios of all transitions except the one to the ground (or isomeric) state which has no subsequent $\gamma$ ray.  However, the branching
ratio of the latter transition can be obtained by subtracting the sum of the former branching ratios from 100\%.  In the case of the \nuc{38}{Ca} decay, which is shown in
Fig.\,\ref{fig:38decays}, the ``missing" transition is the superallowed one, and the subtraction from 100\% actually has a very salutary effect on the relative precision,
reducing it by more than a factor of 3 (= 22.7/77.3).  Very recently with this system, the branching ratio for the \nuc{38}{Ca} superallowed transition has been determined
to $\pm$0.2\% \cite{Pa14}.  This completes the information needed to obtain a precise $ft$ value for this transition and will add \nuc{38}{Ca} to the current list of well known
superallowed decays (see Fig.\,\ref{fig:chart}), the first to be added in nearly a decade.  There will likely be several more additions of this type in the next few years.

\subsection{\label{survey} Survey of world data}

Many independent measurements contribute to the determination of superallowed $ft$ values, so for the past four decades we have periodically produced critical surveys
of relevant world data available at the time of writing.  All published measurements are carefully considered, with some being rejected but only if a specific fault
has been identified.  Of the surviving results, only those with uncertainties that are within a factor of 10 of the most precise measurement for each quantity are retained.
They are then averaged by the same procedures as those adopted by the Particle Data Group \cite{PDG12}.  In columns two, three and four of table~\ref{Ftdata} we present the
results from the most recent survey \cite{Ha09}. In almost all cases, the tabulated values are averages of several -- sometimes many -- experimental measurements
with comparable uncertainties.

For each superallowed transition, the three measured quantities, $Q_{EC}$, $t_{1/2}$ and $R$, together with a small correction ($\sim$0.1\%) \cite{Ha09} to account for
the contribution of electron capture, are combined to obtain an $ft$ value.  These $ft$ values appear in column five of the table.  The next step is to use (\ref{Ftconst})
to obtain the corrected $\F t$ value for each transition.  Before doing this though, we need to examine the theoretical correction terms used to account for radiative and
isospin-symmetry-breaking effects. 

\begin{table}[t]
\begin{center}
\caption{\label{Ftdata}Experimental results ($Q_{EC}, t_{1/2}$ and $R$) and calculated
corrections terms discussed in the text ($\delta_R^{\prime}$, $\delta_{NS}$ and $\delta_C$) for the precisely
measured $0^+$$\rightarrow$~$0^+$ transitions, together with their derived $ft$ and $\F t$
values \cite{Ha09}.}
\lineup
{\scriptsize
\begin{tabular}{@{}lllllllll}
\br
\0 Parent & $Q_{EC}$ & $t_{1/2}$ & $R$ & $ft$ & $\delta_R^{\prime}$
 & $\delta_{NS}$ & $\delta_C$ & $\F t$ \\
\0 nucleus & (keV) & ($ms$) & (\%) & ($s$) & (\%) & (\%) & (\%) & ($s$) \\
\mr
\0   $^{10}$C  & 1907.87(11) & 19308(4) &  1.4646(19) & 3042(4)
   & 1.679(4)& $-0.35(4)$ & 0.175(18) & 3076.7(46)\\ 
\0   $^{14}$O  & 2831.24(23) & 70620(15) &  99.37(7) & 3042.3(27)
   & 1.543(8) & $-0.25(5)$ & 0.330(25) & 3071.5(33)\\   
\0   $^{22}$Mg & 4124.55(28) & 3875.2(24) &  53.16(12) &  3052(7)
   & 1.466(17) & $-0.225(20)$ & 0.380(22) & 3078.0(74)\\ 
\0   $^{26}$Al$^m$ & 4232.66(6) & 6345.0(19) &  $>$99.997 & 3036.9(9)
   & 1.478(20) & \m 0.005(20) & 0.310(18) & 3072.4(14)\\ 
\0   $^{34}$Cl & 5491.64(23) & 1526.6(4) &  $>$99.988 & 3049.4(12)
   & 1.44(3) & $-0.085(15)$ & 0.65(5) & 3070.6(21)\\ 
\0   $^{34}$Ar & 6063.0(5) & 843.8(4) &  94.45(25) & 3053(8)
   & 1.423(29) & $-0.180(15)$ & 0.67(6) & 3069.6(85)\\ 
\0   $^{38}$K$^m$ & 6044.40(11) & 924.33(27) &  99.967(4) & 5051.9(10)
   & 1.44(4) & $-0.100(15)$ & 0.66(6) & 3072.5(24)\\ 
\0   $^{42}$Sc & 6426.3(3) & 680.72(26) &  99.9941(14) & 3047.6(14)
   & 1.45(5) & \m 0.035(20) & 0.67(6) & 3072.4(27)\\ 
\0   $^{46}$V  & 7052.49(16) & 422.59(11) &  99.985($^{+1}_{-4}$) & 3049.5(9)
   & 1.45(5) & $-0.035(10)$ & 0.62(6) & 3073.3(27)\\ 
\0   $^{50}$Mn & 7634.45(7)  & 283.21(11) &  99.942(3) & 3048.4(12)
   & 1.44(6) & $-0.040(10)$ & 0.66(5) & 3070.9(28)\\ 
\0   $^{54}$Co & 8244.37(28) & 193.27(6) &  99.996($^{+1}_{-30}$) & 3050.8($^{+11}_{-15}$)
   & 1.44(7) & $-0.035(10)$ & 0.77(7) & 3069.9(33)\\ 
\0   $^{62}$Ga & 9181.1(5) & 116.12(4) &  99.862(11) & 3074.1(15)
   & 1.46(9) & $ -0.045(20)$ & 1.48(21) & 3071.5(72)\\
\0   $^{74}$Rb & 10417(4) & 64.78(4) &  99.50(10) & 3085(8) 
   & 1.50(12) & $-0.075(30)$ & 1.6(3) & 3078(13)\\[3mm]
   & & & & & & \multicolumn{2}{r}{Average, $\overline{\F t}$} & $3072.08(79)$ \\
   & & & & & & \multicolumn{2}{r}{$\chi^2/\nu$} & \multicolumn{1}{c}{0.28} \\
   
\br
\end{tabular}
}
\end{center}
\end{table}

\section{\label{s:theo} Theoretical corrections}

\subsection{\label{ss:radc} Radiative corrections}

As described in the Introduction, the historical determination of the
vector coupling constant $\GV$ in the semi-leptonic decays of
\nuc{14}{O} and \nuc{10}{C} differed by a few percent from $\GF$, which
was obtained from the purely leptonic decay of the muon -- seemingly 
violating universality as espoused by Cabibbo.  At the time though
it was understood that radiative corrections could largely explain 
the discrepancy.  Simply put, the lifetimes measured in the decays of
\nuc{14}{O} and \nuc{10}{C} include both the bare $\beta$-decay process
and the radiative process, in which the emitted electron releases
a bremsstrahlung photon that is undetected.  Since $\GV$ is
determined from the bare $\beta$-decay process, the contribution from
radiative effects needs to be computed and subtracted from
the measured result.  Calculations of this contribution, however, have to
consider not just the bremsstrahlung process alone but also the
loop diagram in which a photon is exchanged between charged particles.
The key point is that the bremsstrahlung contribution diverges in the
limit when the photon energy goes to zero; the loop graph likewise 
diverges but with the opposite sign, so the combination of the two
remains finite in the low-energy limit.  That this happens is a
consequence of the renormalizability of quantum electrodynamics.

There was another problem for the calculations performed
in the 1950s \cite{Be58,KS59}: the loop graph also diverged in the 
limit of infinite photon energy, so some form of arbitrary cut-off 
had to be imposed.  The difficulty at the time was what to choose for the cut-off: As the cut-off
increased so did the discrepancy between $\GV$ and $\GF$.
In this era the $\beta$-decay process was treated as a four-fermion
contact interaction and the loop diagram was of triangular geometry.

During the 1960s, the Standard Model began to be formulated and the
four-fermion contact interaction was replaced by one in which
an intermediate vector boson -- the $W$ boson -- mediated between
the leptons and hadrons in semi-leptonic decay.
With the $W$ boson, the loop diagram became of rectangular geometry
and the mass of the boson then provided a natural cut-off \cite{Si67}.
The Standard Model also introduced a neutral $Z$ boson
that mixed with the photon, so a further class of loop diagrams
involving $Z$-boson exchange had to be included in the
radiative-correction calculation.  The main effect of this addition
was to increase the effective cut-off in the loop diagram from
the $W$-boson mass to the $Z$-boson mass \cite{Si74,Si78}. 

It was realized all along that the purely leptonic muon decay was also
subject to radiative corrections and that many of the potential
contributions were identical in both pure- and semi-leptonic decays.
Such identical contributions are called `universal'.  Thus, in the 
Standard Model, $\GF$ came to be defined as the weak-interaction
coupling constant, with the understanding that it included within it all universal radiative
corrections.  Thus, the radiative correction applied to the
semi-leptonic decays only needed to include terms that were non-universal.
A longer discussion of this point is given by Sirlin \cite{Si74}.

So far, the radiative correction was only calculated to lowest order in the 
fine-structure constant $\alpha$.  In the 1970s, Jaus and Rasche
\cite{JR70,Ja72} gave the first estimate of the order-$Z \alpha^2$
contributions, where $Z$ is the charge number of the daughter nucleus.  
This correction is defined as the contribution at this order not already
contained in the product $F(Z,E) (1 + \delta_1)$, where $F(Z,E)$
is the Fermi function, $E$ the electon total energy and $\delta_1$
the order-$\alpha$ correction.  Also in the 1970s, experimental 
results of superallowed beta decay began to emerge in the higher-$Z$
elements of \nuc{42}{Sc}, \nuc{46}{V}, \nuc{50}{Mn} and \nuc{54}{Co}.
These results raised a problem \cite{TH84}: the high-$Z$ cases seemed
to be failing the required Cabibbo universality.  This prompted
a reexamination of the order-$Z \alpha^2$ radiative correction
by Sirlin and Zucchini \cite{SZ86, Si87} and Jaus and Rasche \cite{JR87}.
An error was discovered in the earlier work and Cabibbo universality
was duly recovered.

In the loop graph there is another interesting wrinkle.  If the nucleus
is treated as a collection of nucleons, and if the weak and the
electromagnetic interactions occurring in the radiative correction
interact with the same nucleon, then the graph evaluation leads to a result 
that is proportional to the expectation value of the isospin ladder 
operator.  Since the bare $\beta$-decay process also proceeds via
the isospin ladder operator, the radiative correction simply scales with
the bare $\beta$-decay value.  This implies that the calculation is
independent of nuclear structure, the result depending only on the
electron energy and the charge $Z$ of the daughter nucleus.
If, however, the weak and electromagnetic interactions occur with
different nucleons, then the scaling property fails
and a full nuclear-structure-dependent calculation is required.
The first such calculation was provided by Jaus and Rasche \cite{JR90,
BBJR92} and later refined by Towner \cite{To92,To94}.
These results constitute the $\delta_{NS}$ correction, which appears in (\ref{Ftconst}).

Another property of the loop graph is that one can usefully employ
different approximations for the hadronic structure depending upon whether
the photon energy is small or large.  At low photon energy, it is
sufficient to treat the nucleus as a collection of nucleons and
use nucleon weak and electromagnetic form factors at the vertices.
At high photon energy, the hadronic structure is essentially that
of a soup of quarks.  In this limit, the calculation becomes independent 
of the details of hadronic structure and leads to the term
$\DRV$ in (\ref{Ftconst}).  Marciano and Sirlin \cite{MS06}
have critically examined this separation into low- and high-energy
contributions and the linkages between them.  Their recommended
value for $\DRV$ is in current use:
\be
\DRV = (2.361 \pm 0.038) \% .
\label{DRV}
\ee
Lastly, a correction to order $\alpha^2$ was recently considered by
Czarnecki, Marciano and Sirlin \cite{CMS04} and is now included
in current computations.

All contributions to the radiative
correction considered to date are collated into $\delta_R^{\prime}$,
$\delta_{NS}$ and $\DRV$ as displayed in (\ref{Ftconst}).
The currently accepted values \cite{TH08} for $\delta_R^{\prime}$ and
$\delta_{NS}$ are given in columns six and seven of table~\ref{Ftdata}.

\subsection{\label{ss:deltc} Isospin symmetry-breaking corrections}

The Conserved Vector Current (CVC) hypothesis asserts that the vector
coupling constant $\GV$ is not renormalized in the nuclear medium
but is the same for all nuclei.  This assertion can be tested by
the measurement of superallowed $\beta$ decays, which cover a wide range of
nuclei spanning from \nuc{10}{C} to \nuc{74}{Rb}.  However, the CVC
hypothesis is only valid in the isospin-symmetry limit, so if we
are to determine $\GV$ from a wide range of nuclei then a correction
has first to be applied that removes the effects of isospin-symmetry breaking.
Isospin symmetry is naturally broken in nuclei because
the protons are subject to the Coulomb interaction, which is not felt by the neutrons.
Thus, the wave function of a proton in a given quantum state differs 
slightly from a neutron in the mirror nucleus in the same quantum
state.  This is reflected in the nuclear matrix element describing
the beta-decay transition between mirror states.  The isospin
symmetry-breaking correction $\delta_C$ is defined by the relation
\be
|M_F|^2 = |M_F^0|^2 (1 - \delta_C ) ,
\label{MFdc}
\ee
where $M_F$ is the exact Fermi matrix element and $M_F^0$ its
isospin-symmetry-limit value, being just the expectation value of
the isospin ladder operator. 

One of the earliest estimates of $\delta_C$ was provided by 
Damgaard \cite{Da69} in 1969.  He suggested that the radial function of
the proton in beta decay could be expanded in a complete set of
neutron radial functions of the same angular momentum; the terms
in the set differing in the number of radial nodes in the function.
The expansion coefficients were determined in perturbation theory,
with the Coulomb force being the perturbing interaction.  With the
basis states taken as harmonic oscillator functions, Damgaard
obtained the expression
\be
\delta_C = \frac{Z^2}{(\hbar \omega )^4 r^6} \frac{e^2 \hbar^4}{16 m^2}
(n + 1)(n + \ell + \sfrac{3}{2}) .
\label{dam1}
\ee
If one adopts the relationships for the characteristic oscillator energy,
$\hbar \omega = 41 A^{-1/3}$ MeV, and the nuclear radius, $r = 1.2 A^{1/3}$
fm,  this expression becomes
\be
\delta_C = 0.2645 Z^2 A^{-2/3} (n+1) (n + \ell +\sfrac{3}{2}) ,
\label{dam2}
\ee
which, for the relatively light nuclei we are interested in here, exhibits the general
behaviour $\delta_C \propto Z^{4/3}$ with some shell structure 
superimposed through the choice of oscillator quantum numbers $n$
and $\ell$.  In particular, a proton radial function with one
radial node gets a factor of 2 enhancement in its $\delta_C$ value 
over one that has no radial nodes simply from the factor $(n+1)$
in (\ref{dam2}).  This enhancement is clearly evident in table~\ref{Ftdata}
for the upper $pf$ shell, where there is a significant increase in experimental
$ft$ value when going from \nuc{54}{Co} ($ft$=3051\,s) to \nuc{62}{Ga} ($ft$=3074\,s).

Beginning in 1973, Damgaard's approach has been improved upon by Towner and
Hardy \cite{To73,TH08,To77,To02} in two significant ways.  The
oscillator basis states have been replaced by eigenfunctions of a
Woods-Saxon potential and the extreme non-interacting single-particle
model has been replaced by the interacting shell model.  The perturbing 
Coulomb force is long range and influences a huge number of configurations
in the shell model.  Thus it has become expedient to divide
$\delta_C$ into two components
\be
\delta_C = \delta_{C1} + \delta_{C2}~,
\label{dc1dc2}
\ee
where $\delta_{C1}$ is a contribution from a finite-sized shell-model
calculation, typically containing all configurations within one
major shell, while $\delta_{C2}$ steps outside that model space
to states that typically have a different number of radial nodes
influencing the proton and neutron radial functions.  For superallowed transitions between
the $0^+$ states of interest here, Towner and Hardy \cite{TH08} find
that the $\delta_{C1}$ contributions are much smaller than the $\delta_{C2}$ ones.
Their results are shown in table~\ref{t:tab2}, where they are labelled SM-WS.
We note that these results are considered to be semi-phenomenological since
a number of isospin-specific nuclear properties have been fitted in their
derivation: $viz.$~the measured proton and neutron separation
energies in the parents and daughters, respectively, were used in the calculation
of $\delta_{C2}$; and each calculation of $\delta_{C1}$ was tuned to fit the
$b$- and $c$-coefficients in the isobaric multiplet mass equation corresponding
to the specific $T=1$ multiplet that included the parent and daughter state.

From 1985 to 1995, Ormand and Brown \cite{OB85,OB89,OB95} adopted the
same general procedure as the one just described but used eigenfunctions
of a Hartree-Fock mean field rather than a Woods-Saxon potential.
Their results were systematically smaller than the SM-WS values and
this difference was used to assess a systematic error in the analysis
of superallowed beta-decay data.  This systematic difference, however,
was much reduced by a 2009 calculation of Hardy and Towner \cite{Ha09},
in which the protocol for the Hartree-Fock calculation was altered to
ensure that the Coulomb part of the proton mean field obtained from 
the Hartree-Fock calculation had the appropriate asymptotic form.
The results from this calculation are labelled SM-HF in table~\ref{t:tab2}.

Besides Damgaard's model and the two shell-model approaches to $\delta_C$,
which apply to the full range of measured superallowed transitions, there have
been many other less-complete computations by various 
authors \cite{Sa86,Li09,Au09,MS08,MS09,Gr10,Sa12} using a diverse 
set of nuclear models.  There are too many to discuss all these cases
here.  We have selected two of them, one based on the random phase approximation (RPA)
and the other on density functional theories (DFT) to include in table~\ref{t:tab2}.
In the former, the RPA work of Sagawa {\em et al} \cite{Sa86}, improved upon
by Liang {\em et al} \cite{Li09}, treats the even-even nucleus of the
parent-daughter pair as a core, and the analogue odd-odd nucleus as
a particle-hole excitation built on that core.  The particle-hole
calculation is carried out in the charge-exchange version of the RPA.
The more recent of the two works \cite{Li09} replaces zero-range interactions
with finite-range meson-exchange potentials, and a relativistic
rather than nonrelativistic treatment (RHF-RPA) is used.  In a variation of
this approach, density-dependent meson-exchange vertices were introduced
in a Hartree (only) computation with nonlocal interactions (RH-RPA).  Both these
sets of results are listed in table~\ref{t:tab2}.

Most recently, Satula {\em et al} \cite{Sa12} used an isospin- and
angular-momentum-projected density functional theory (DFT).  This
method accounts for spontaneous symmetry breaking, configuration
mixing and long-range Coulomb polarization effects.  The results are also
listed in table~\ref{t:tab2}.

\begin{table}
\caption{\label{t:tab2}Six sets of $\delta_C$ calculations from model       
approaches labelled Damgaard, DFT, RHF-RPA, RH-RPA, SM-HF and SM-WS
(see text).  Also given is the chi square per degree of freedom,
$\chi^2/n_d$, from the confidence test proposed in ref.~\cite{TH10}.}
\lineup
\begin{indented}
\item[]\begin{tabular}{@{}lllllll}
\br
& \multicolumn{6}{c}{$\delta_C(\%)$} \\[-1mm]
& \crule{6} \\
\0 Nucleus &
Damgaard &
DFT &
RHF-RPA &
RH-RPA &
SM-HF &
SM-WS \\
\mr
\0 \nuc{10}{C}  &  0.046 & 0.462(65) & 0.082 & 0.150 & 0.225(36) & 0.175(18)  \\
\0 \nuc{14}{O}  &  0.111 & 0.480(48) & 0.114 & 0.197 & 0.310(36) & 0.330(25)  \\
\0 \nuc{22}{Mg} &  0.153 & 0.432(49) &       &       & 0.260(56) & 0.380(22)  \\
\0 \nucm{26}{Al}&  0.182 & 0.307(62) & 0.139 & 0.198 & 0.440(51) & 0.310(18)  \\
\0 \nuc{34}{Cl} &  0.326 &           & 0.234 & 0.307 & 0.695(56) & 0.650(46)  \\
\0 \nuc{34}{Ar} &  0.285 & 1.08(42)  & 0.268 & 0.376 & 0.540(61) & 0.665(56)  \\
\0 \nucm{38}{K} &  0.370 &           & 0.278 & 0.371 & 0.745(63) & 0.655(59)  \\
\0 \nuc{42}{Sc} &  0.414 & 0.70(32)  & 0.333 & 0.448 & 0.640(56) & 0.665(56)  \\
\0 \nuc{46}{V}  &  0.524 & 0.375(96) &       &       & 0.600(63) & 0.620(63)  \\
\0 \nuc{50}{Mn} &  0.550 & 0.39(13)  &       &       & 0.620(59) & 0.655(54)  \\
\0 \nuc{54}{Co} &  0.613 & 0.51(20)  & 0.319 & 0.393 & 0.685(63) & 0.770(67)  \\
\0 \nuc{62}{Ga} &  1.34  &           &       &       & 1.21(17)  & 1.48(21)   \\
\0 \nuc{74}{Rb} &  1.42  & 0.90(22)  & 1.088 & 1.258 & 1.42(17)  & 1.63(31)  \\[2mm]
\0 $\chi^2 / n_d$ & 1.7 & 1.9 & 2.7 & 2.1 & 2.2 & 0.4 \\
\br
\end{tabular}
\end{indented}
\end{table}

The six sets of $\delta_C$ values in table~\ref{t:tab2} show a wide 
variation.  Some yardstick is required to distinguish the quality of one
set relative to another, so Towner and Hardy \cite{TH10} proposed such a
test using the premise
that the CVC hypothesis is valid.  The requirement is that a
calculated set of $\delta_C$ values should produce a statistically
consistent set of $\F t$ values, the average of which we can
write as $\overline{\F t}$.  Then (\ref{Ftconst}) can be written
for each individual transition in the set as
\be
\delta_C = 1 + \delta_{NS} - \frac{\overline{\F t}}{ft (1 + \delta_R^{\prime})}.\label{dctest}
\label{deltaCtest}
\ee
For any set of corrections to be acceptable, the calculated value of
$\delta_C$ for each superallowed transition must satisfy this
equation, where $ft$ is the measured result for the transition and
$\overline{\F t}$ has the same value for all of them.  Thus, to test a
set of $\delta_C$ values for $n$ superallowed transitions, one can
treat $\overline{\F t}$ as a single adjustable parameter and use it
to bring the $n$ results for the right-hand side of (\ref{dctest}),
which are based predominantly on experiment, into the best
possible agreement with the corresponding $n$ calculated values
of $\delta_C$ on the left-hand side of the equation.  The normalized $\chi^2$, minimized by this process
then provides a figure of merit for that set of calculations.
The $\chi^2$ for each fit, expressed as $\chi^2/n_d$, where
$n_d = n-1$ is the number of degrees of freedom, is given in the
last row of table~\ref{t:tab2}.

The most obvious outcome of this analysis is that the model, SM-WS,
has a $\chi^2$ smaller by a factor of five than the other five cases cited.  
For this reason, the SM-WS $\delta_C$ values are the ones used in the calculation of
the $\F t$ values, and it is these $\delta_C$ values that appear in the eighth column of table~\ref{Ftdata}.
However, the other $\delta_C$ calculations can be used to help establish a systematic-uncertainty
assignment on this analysis.

\section{\label{s:impact} Impact on weak-interaction physics}

\subsection{\label{ss:Vud} The value of $V_{ud}$}
  
The $\F t$ values obtained with the Woods-Saxon 
isospin-symmetry breaking correction SM-WS appear in the ninth column of table~\ref{Ftdata}, where
all are seen to be mutually consistent.  Thus, we are justified in averaging the 13
entries in table~\ref{Ftdata} and using the result, $\overline{\F t} = 3072.08 \pm
0.79$~s, to determine the CKM matrix element $V_{ud}$.  However, 
before doing so we must consider the impact that a
different set of isospin-symmetry breaking corrections might have on the
result.  In the past \cite{Ha05}, we
compared the SM-WS calculations with the Hartree-Fock calculations
of Ormand and Brown \cite{OB85,OB89,OB95}, whose $\delta_C$ corrections
covered all measured transitions and were consistently smaller than those obtained with Woods-Saxon functions.
We considered that this provided a valid estimate of the systematic (theoretical) uncertainty, 
and so incorporated it into the overall result by deriving two average $\overline{\F t}$
values, one for each set of $\delta_C$ calculations, then adopting the
average of the two and assigning a systematic uncertainty equal to 
half the spread between them.

This procedure was continued in the most recent 2009 survey \cite{Ha09},
except that the Ormand and Brown calculations were replaced by the Towner-Hardy
Hartree-Fock values, SM-HF in table~\ref{t:tab2}.  With the SM-HF $\delta_C$
values, the average $\overline{\F t}$ value became $3071.55 \pm 0.89$~s but with
a substantially increased chi-square of $\chi^2/\nu = 0.93$.  This normalized chi-square is
three times what was obtained in table~\ref{Ftdata} with Woods-Saxon
corrections, which arguably could justify a rejection of the
Hartree-Fock results outright.  However, to be safe, we proceeded as in
the 2005 survey and take the average of the SM-WS and SM-HF results,
adding a systematic uncertainty equal to half the spread between the
two results.  Thus, the 2009 survey \cite{Ha09} arrived at
\bea
\overline{\F t} & = & 3071.81 \pm 0.79_{\rm stat} \pm 0.27_{\rm syst}~{\rm s}
\nonumber \\
& = & 3071.81 \pm 0.83~{\rm s}~,
\label{Ftfinal}
\eea  
where on the second line the two uncertainties have been added in quadrature.

With $\overline{\F t}$ thus obtained, the CKM matrix element $V_{ud}$ is derived from
a rearrangement of (\ref{GVtoVud}) and (\ref{Ftconst}):
\be
|V_{ud}|^2 = \frac{K}{2 \GF^2 (1+\DRV ) \overline{\F t}}
= \frac{2915.64 \pm 1.08}{\overline{\F t}} ,
\label{Vud2}
\ee
where the overall weak-interaction coupling constant from muon decay is
$\GF /(\hbar c)^3 = 1.1663787(6) \times 10^{-5}$~GeV$^{-2}$ from
Ref.\,\cite{PDG12} and $\DRV$ is taken from (\ref{DRV}).  On applying our
recommended value of $\overline{\F t}$ from (\ref{Ftfinal}) we arrive at
\be
|V_{ud}| = 0.97425 \pm 0.00022 ,
\label{Vud}
\ee
a value with $0.02 \%$ precision.

This result is certainly the most precise current determination of $V_{ud}$,
but superallowed $0^+ \rightarrow 0^+$ $\beta$ decay is not the only 
experimental approach to $V_{ud}$.  Neutron decay, nuclear $T = 1/2$ mirror
decays and pion beta decay have all been used for this purpose.  For now,  
these other methods cannot compete with $0^+ \rightarrow 0^+$
decays for precision, although they yield statistically consistent results \cite{TH10}.
We need not consider them farther in this context.

\subsection{\label{ss:unitarity} Unitarity of the CKM matrix}

That the sum of the squares of the top-row elements should add to one
is the most stringent test of the CKM matrix's unitarity.  Here $V_{ud}$
plays the dominant role with by far the largest magnitude and the smallest relative
uncertainty; but, when it comes to the unitarity sum, $V_{us}$
and $V_{ud}$ contribute equally to the uncertainty because the terms themselves have
such different magnitudes.  For $V_{us}$ we will
use the value reported at the recent CIPANP 2012 conference \cite{Mo12} (which
updates the 2012 Particle Data Group value \cite{PDG12}):
\be
|V_{us}| = 0.2256 \pm 0.0008 .
\label{Vus}
\ee
This value is an average of results obtained from
kaon semi-leptonic decays, $K_{\ell 3}$, of both charged and neutral kaons,
and from the purely leptonic decay of the kaon, $K_{\ell 2}$.  Both
methods rely on lattice QCD calculations for values of the hadronic
form factors.  Other determinations from hyperon decays and
hadronic tau decay do not have the precision at the present time
to challenge the results from kaon decays.

The third element of the top row of the CKM matrix, $V_{ub}$, is very
small and hardly impacts on the unitarity test at all.  Its value
from the 2012 PDG compilation \cite{PDG12} is
\be
|V_{ub}| = (4.15 \pm 0.49) \times 10^{-3} .
\label{Vub}
\ee

Combining the values given in (\ref{Vud}), (\ref{Vus}) and (\ref{Vub}), the
sum of the squares of the top-row elements of the CKM matrix becomes
\bea
|V_{ud}|^2 + |V_{us}|^2 + |V_{ub}|^2 & = &
1.00008 \pm 0.00043_{V_{ud}} \pm 0.00036_{V_{us}} , 
\nonumber \\
& = & 1.00008 \pm 0.00056 ,
\label{Usum}
\eea
a result that shows unitarity to be fully satisfied at the $0.06 \%$ level.
In the first line of (\ref{Usum}), the two errors shown are firstly
from the uncertainty in $V_{ud}$ and secondly
from $V_{us}$.  They are combined in quadrature in the second line. 
Observe that kaon decays contribute slightly less than nuclear decays to the
error budget.

No other row or column approaches this precision on a unitarity test.  The 
first column comes closest, with
$|V_{ud}|^2+|V_{cd}|^2+|V_{td}|^2 = 1.0021 \pm 0.0051$ \cite{CLS12},
but this is a factor of ten less precise than the top-row sum.
The corresponding sums for the second row and second column are $1.067 \pm 0.047$
and $1.065 \pm 0.046$ respectively \cite{CLS12}, another order of magnitude
less precise.  Without question the top-row sum provides the 
most demanding test of CKM unitarity, $V_{ud}$ is its dominant
contributor, and superallowed $\beta$ decay is effectively the
sole experimental source for the value of $V_{ud}$.

The excellent experimental agreement with unitarity provides strong
confirmation of the Standard-Model radiative corrections that enter both
nuclear and kaon decays at the 3 to $4 \%$ level and, to a lesser extent,
confirmation of isospin-symmetry breaking estimates, again in
both nuclear and kaon decays.  In addition it implies constraints on
new physics beyond the Standard Model.  New physics can enter in one of two 
ways: directly, via a new semi-leptonic interaction ({\em e.g.} scalar currents or right-hand
currents), or indirectly, via loop-graph contributions to the
radiative corrections ({\em e.g.} extra $Z$ bosons).  Discussions of
these issues can be found in \cite{TH10,BM12,NG13} as well as in other
contributions to this volume.

\section{\label{s:future} Future Prospects}

Because the unitary CKM matrix is an essential pillar of the Standard Model, the uncertainty limits
on the unitarity sum in (\ref{Usum}) constrain the scope of whatever new physics may be anticipated to lie beyond that model.
Consequently, there is ample motivation to search for improvements in both theory and experiment that can
reduce the uncertainty on the sum, and potentially expose -- or rule out -- some classes of
new physics.

Currently the uncertainty in the value of $V_{us}$ is dominated by the lattice QCD estimate for the
form factor used to extract it from the experimental measurements on $K_{\ell 3}$ decays.  It is predicted,
though, that these lattice calculations will improve considerably over the next decade, with reasonable prospects
of reducing the $V_{us}$ uncertainty by a factor of 2 \cite{Mo12, Ha13}.  With this improvement in sight, it
is evident from (\ref{Usum}) that any future improvements in $V_{ud}$ will thus have a significant impact on
the overall uncertainty of the unitarity sum.  Are such improvements foreseeable?

It is the uncertainties on the calculated correction terms, particularly $\DRV$, $\delta_C$ and $\delta_{NS}$,
that have the greatest influence on the uncertainty of $V_{ud}$.  The largest contributor, $\DRV$, may be open
to some improvement, with a 30\% reduction in uncertainty having been suggested as possible \cite{Ma13}.  This
is a challenge exclusively for theorists.  However, improvements in the nuclear-structure-dependent corrections, $\delta_C$
and $\delta_{NS}$, can actually be achieved with the help of experiments.  As described in Section \ref{ss:deltc}, these
terms are subject to a test: They can be applied to the uncorrected experimental $ft$ values to obtain a set
of $\F t$ values, which can then be evaluated for the consistency required by CVC -- see (\ref{deltaCtest}) and
table~\ref{t:tab2}.  The more precisely the $ft$ values have been measured, the more demanding this test can be; and
if new superallowed transitions with larger predicted nuclear corrections can be measured, the test will be improved
still more.

Of particular importance in this context are superallowed $0^+$$\rightarrow 0^+$\,decays of the $T_Z$\,=\,$-1$ nuclei
$^{26}$Si, $^{34}$Ar, $^{38}$Ca and $^{42}$Ti.  As noted in Section \ref{TZ-1}, these decays are more complex than the
currently well studied superallowed transitions, which mostly have $T_Z$\,=\,0 parents.  Only very recently have $T_Z$\,=\,$-1$ decays
in this mass region become amenable to precise $ft$-value measurements \cite{Pa14} and it is anticipated that all four will be
fully characterized within the next few years.  These cases will be influential not only because they have relatively large
nuclear corrections but also because each of the four transitions is mirror to another well known transition, $^{26m}$Al, $^{34}$Cl,
$^{38m}$K and $^{42}$Sc, respectively.  It turns out that the ratio of the $ft$-values for mirror superallowed transitions is
extremely sensitive to the model used to calculate $\delta_C$ and $\delta_{NS}$ \cite{Pa14}.  There is good reason to expect that
these new cases will tighten the model constraints enough to shrink or even remove entirely the systematic uncertainty now
applied to the structure-dependent correction terms. 

The potential improvements in all the correction terms should act to reduce the uncertainty in $V_{ud}$ by about 25\% and,
together with the improvements expected in $V_{us}$, can be expected to reduce the uncertainty in the CKM unitarity sum --
see (\ref{Usum}) -- from $\pm$0.00056 to $\pm$0.00037.  This is likely to be the extent of possible improvements in the foreseeable
future.

\ack

This work was supported by the U.S. Department of Energy under Grant No.\,DE-FG03-93ER40773 and by the Robert
A. Welch Foundation under Grant No.\,A-1397.

\end{document}